\documentclass[a4paper,twocolumn]{revtex4-1}
\usepackage[english]{babel}
\usepackage[utf8]{inputenc}
\usepackage{amsmath}
\usepackage{amsfonts}

\usepackage{bm}
\usepackage{siunitx}

\newcommand{\ket}[1]{\vert #1 \rangle}
\newcommand{\op}[2]{\vert #1 \rangle\langle #2 \vert}

\newcommand{\mel}[3]{\langle #1 \vert #2 \vert #3 \rangle}

\newcommand{\tr}[1]{\mathrm{Tr}\left(#1\right)}
\newcommand{\diag}[1]{\mathrm{diag}(#1)}
\newcommand{\p}{\partial}

\newcommand{\mat}[1]{\begin{pmatrix} #1 \end{pmatrix}}

\usepackage[colorlinks=true,linkcolor=black,citecolor=black]{hyperref}
\usepackage[capitalise]{cleveref}

\usepackage{graphicx}

\usepackage{natbib}

\newcommand{\affA}{Department of Physics and Astronomy, Aarhus University, DK-8000 Aarhus C, Denmark}
\newcommand{\affB}{Aarhus Institute of Advanced Studies, Aarhus University, DK-8000 Aarhus C, Denmark}

\date{\today}
\begin{document}
	
	\title{Native three-body interaction in superconducting circuits}
	
	\author{Simon Panyella Pedersen}
	\email{simon.panyella.pedersen@post.au.dk}
	\affiliation{\affA}
	\author{K. S. Christensen}
	\affiliation{\affA}
	\author{N. T. Zinner}
	\email{zinner@phys.au.dk}
	\affiliation{\affA}
	\affiliation{\affB}

\begin{abstract}
	We show how a superconducting circuit consisting of three identical, non-linear oscillators in series considered in terms of its electrical modes can implement a strong, native three-body interaction among qubits. Because of strong interactions, part of the qubit-subspace is coupled to higher levels. The remaining qubit states can be used to implement a restricted Fredkin gate, which in turn implements a CNOT-gate or a spin transistor. Including non-symmetric contributions from couplings to ground and external control we alter the circuit slightly to compensate, and find average fidelities for our implementation of the above gates above $ 99.5\% $ with operation times on the order of a nanosecond.	Additionally we show how to analytically include all orders of the cosine contributions from Josephson junctions to the Hamiltonian of a superconducting circuit. 
\end{abstract}

\maketitle

\section{Introduction}\label{sec:intro}
Superconducting circuits offer a potent platform for the implementation of quantum simulation and computation \cite{Devoret2013,Nori2014,Nori2017,Wendin2017}. In quantum simulation and computation multi-body interactions among qubit are important especially for analog simulation of lattice gauge theories \cite{Marcos2013,Marcos2014,Hauke2013,Mezzacapo2015,Zohar2015} and quantum annealing \cite{Lechner2015,Leib2016,Chancellor2017,Biamonte2008,Perdomo2008,PerdomoOrtiz2012}. Proposals for such interactions have involved effective interactions from perturbation theory \cite{Marcos2013,Marcos2014,Hauke2013,Lloyd2016}, mutual inductive couplings to a common resonator \cite{Mezzacapo2015,Chancellor2017,Kafri2017,Schondorf2018} and the decomposition of the multi-body interaction to many two-body interactions involving ancilla qubits \cite{Mezzacapo2015,Lechner2015,Leib2016,Biamonte2008,Perdomo2008,PerdomoOrtiz2012}, similar to decomposition of gates in digital simulation \cite{NielsenChuang2010,Figgatt2017}. Furthermore, multi-body interactions have been proposed using the so called perturbative or Hamiltonian gadgets \cite{Kempe2006,Jordan2008,Bravyi2008,Cao2015}. These gadgets induce effective multi-body interactions via perturbation theory and decomposition into many two-body interactions with a considerable overhead of ancilla qubits.

Here we show how the electrical modes of an idealized, symmetric superconducting circuit can implement a native three-body interaction among qubits. We call these three qubits input, output and control. This is similar to how the electrical modes are used to implement desired interactions in Refs. \cite{Kounalakis2018,Roy2017,Roy2018}, and in Ref. \cite{Sameti2017} a Schrieffer-Wolff transformation is used similarly. Examples of quantum computation using three-body interactions can be seen in \cite{Nagaj2010,Feynman1985}. Being native, the couplings are strong and do not require any ancillary qubits. This contrasts perturbative interactions, which by their very nature are generally weak, and decomposition schemes, where ancilla qubits are necessary. Because the couplings are strong in our set up, interactions between the qubit subspace and higher excited states of the anharmonic oscillators are not completely suppressed. Two of the eight qubit states immediately evolve out of the qubit subspace. We therefore effectively work in a six-state Hilbert space. We refer to the degrees of freedom as qubits to give a clearer intuitive picture and underline the intention of the system as a component in quantum technology or a quantum simulator. The remaining six states can be used to implement a CNOT-gate or a spin transistor, where the control is a quantum degree of freedom. Including non-symmetric contributions from couplings to external control, readout and ground, thus making the circuit more realistic, undesired detuning and couplings are induced, and the circuit must be altered slightly to compensate. After this alteration there still remains an undesired XX-coupling, resulting in uncontrolled hopping between input and output. Even so the average fidelity of our implementation of the above gates is nonetheless above $ 99.5\% $ with operation times on the order of a nanosecond.  These extremely fast operation times are also a consequence of the strong couplings. This time scale is comparable to recent experimental results where a two-qubit gate with an operation time of less than a nanosecond was realised with electron spin qubits bound to phosphorous donors in silicon \cite{He2019}.  Our numerical results for a (restricted) three-qubit gate in the context of superconducting circuits are thus comparable to this two-qubit gate in electron spins, which to the best of our knowledge is one of the fastest realised two-qubit gates.

Furthermore, in this paper we show how to include all orders of the cosine contributions from Josephson junctions to the qubit. We do this with two levels for each qubit in the main text but in the Appendix we show how the same method can be used to include more levels. This is useful for analytical study of suppression of couplings to higher levels and for faster simulation of multiple levels.

This article is organized as follows: In \cref{sec:ideal:circHamil} we go through the Hamiltonian for the idealized circuit. We discuss what the resulting gate is in \cref{sec:ideal:gate}. Then in \cref{sec:real:ext} we move on to the more realistic circuit by introducing couplings to external control and ground. The alteration to the circuit and the subsequent Hamiltonian is discussed in \cref{sec:real:circHamil}. In \cref{sec:sim} we go through the simulation of the system. Spin model parameters are presented in \cref{sec:sim:smp}, the state population evolution in \cref{sec:sim:smp}, and calculations of average fidelity is presented in \cref{sec:sim:fid}. Finally in \cref{sec:discu} we discuss results and conclude the paper in \cref{sec:conc}.

\section{Idealized system}\label{sec:idealsystem}
We consider a very symmetric circuit consisting essentially of three transmon \cite{Transmon2007} blocks in series with identical parameters. This circuit is unrealistic in the sense that it is not connected to external control and readout, or to ground. These external couplings induce an asymmetry in the system, which we will compensate for in the next section. Changing basis to the eigenmodes of the capacitance network, the circuit implements three qubits interacting via an XXC-coupling which flips the input and output states conditioned on the state of the control qubit.

\subsection{Circuit \& Hamiltonian}\label{sec:ideal:circHamil}
\begin{figure*}
	\centering
	\includegraphics[width=0.8\textwidth]{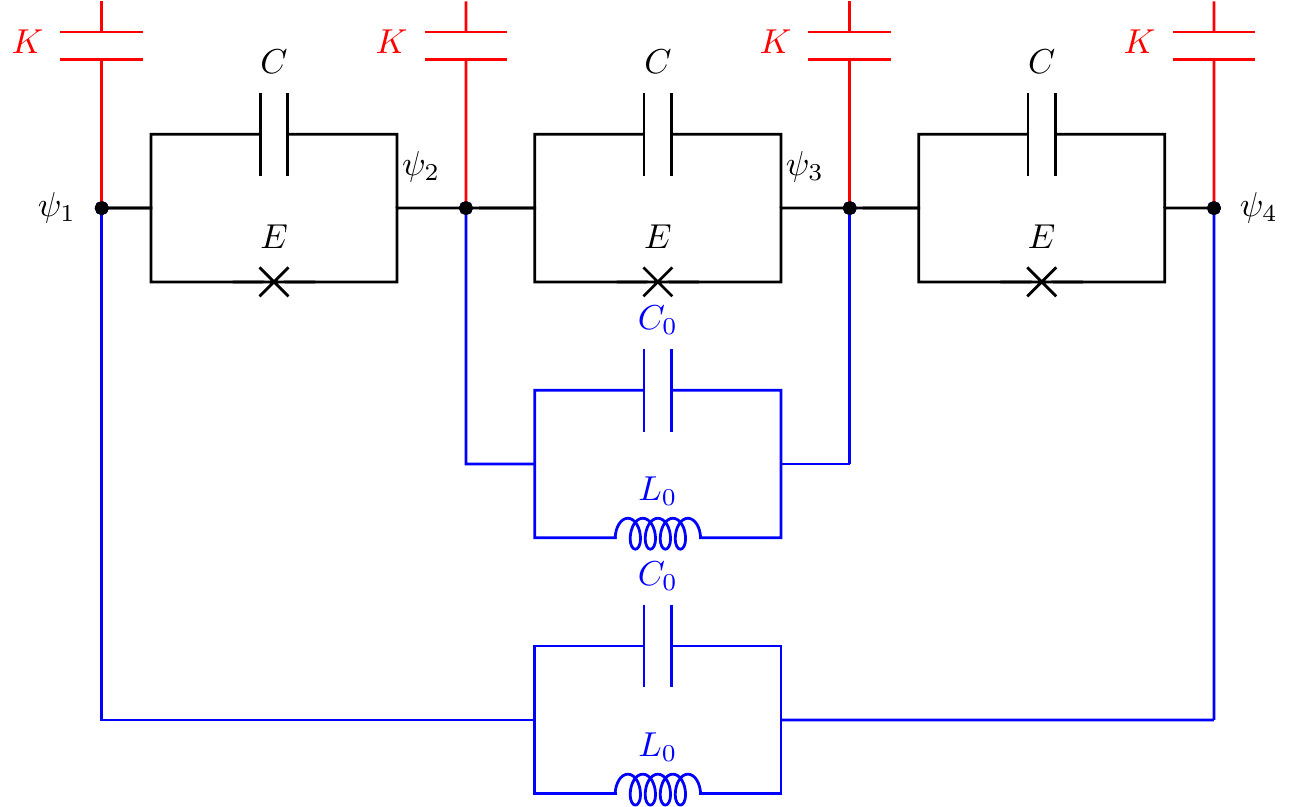}
	\caption{\label{fig:circuit} In black: the idealized, symmetric circuit without external coupling. In red: external couplings to control, readout and ground represented by one joint capacitance for each node. In blue: the additional inductors with parasitic capacitance, introduced to counter the asymmetry of the external couplings.}
\end{figure*}

The circuit we consider can be seen in black in \cref{fig:circuit}. It is similar to that of \cite{Kounalakis2018}. We relate a flux coordinate, $ \psi_{j} $ for $ j = 1,2,3,4 $, to each node in the circuit \cite{Devoret2017}. Gathering these in a vector $ \bm{\psi} $, we change coordinates to the eigenmodes of the capacitance network $ \bm{\phi} = (\phi_{CM},\phi_{1},\phi_{2},\phi_{c})^{T} $, via $ \bm{\psi} = V\bm{\phi} $, where 
\begin{align}
V = \mat{1 & 1+\sqrt{2} & 1-\sqrt{2} & 1 \\ 1 & 1 & 1 & -1 \\ 1 & -1 & -1 & -1 \\ 1 & -1-\sqrt{2} & -1+\sqrt{2} & 1}
\end{align}
is a matrix whose columns are the eigenvectors of the capacitance matrix
\begin{align}
\mathcal{K} = \mat{C & -C & 0 & 0 \\ -C & 2C & -C & 0 \\ 0 & -C & 2C & -C \\ 0 & 0 & -C & C}
\end{align}
In these coordinates the capacitance matrix becomes diagonal, thus avoiding interactions induced by the capacitances. These eigenmodes or electrical modes are associated with circuit charge oscillations between the superconducting islands of the circuit or groups of these \cite{Kounalakis2018}. For example $ \phi_{c} $ corresponds to charge oscillating between the islands of $ \psi_{1} $ and $ \psi_{4} $ and those of $ \psi_{2} $ and $ \psi_{3} $. That is, in this mode charge oscillates between the two ends of circuit and the middle of it. All our modes are quadrupolar as they involve all four of the original coordinates and thus all four of the superconducting islands. In other circuits there might be dipolar modes, involving only two of the original flux coordinates, or higher than fourth order poles, involving more than four coordinates. All but $ \phi_{CM} $ will be anharmonic oscillators and give rise to transmon-like qubits \cite{Transmon2007}. The Hamiltonian in these coordinates is (see \cref{app:idealcirc})
\begin{align}
H^{I} &= \phantom{+} \frac{E_{C}}{2}\left(p_{1}^2 + p_{2}^2 + p_{c}^2\right) \nonumber\\
	&\hspace{11.5pt} - E\cos(2\phi_{1} + 2\phi_{2}) \nonumber\\
	&\hspace{11.5pt} - 2E\cos(\sqrt{2}\phi_{1} - \sqrt{2}\phi_{2})\cos(2\phi_{c}) \label{eq:idealfluxH}
\end{align}
where the $ p_{i} $, $ i = 1,2,c $, are the momenta conjugate to $ \phi_{i} $, and $ E_{C} = 1/8C $ is the charging energy of a capacitor with capacitance $ C $ (in units $ \hbar = 2e = 1 $). The mode $ \phi_{CM} $ does not enter at all and corresponds to a center of mass mode, and so we ignore it from this point on. Recasting the modes in terms of harmonic oscillators and truncating each to its two lowest levels, we can express this Hamiltonian exactly in terms of Pauli matrices. Usually one first expands the cosines to fourth order in the fluxes when working with transmon qubits \cite{Transmon2007}, but by writing the cosines in terms of complex exponentials we can derive the following identity (see \cref{app:idealrecast}), 
\begin{align}
\left(e^{ik(a^{\dagger} + a)}\right)_{2} = \left[\left(1 - \frac{k^2}{2}\right)I + \frac{k^2}{2}\sigma^{z} + ik\sigma^{x}\right]e^{-k^2/2} \label{eq:exptwolevel}
\end{align}
where $ (\cdot)_{2} $ indicates truncation to two levels, $ a^{\dagger} $/$ a $ are bosonic creation/annihilation operators, $ k $ is a real number, and $ \sigma^{x,y,z} $ are the Pauli matrices. For a four level version of this identity see \cref{app:fourlevel}. With this identity the contributions from the cosines to the two-level Hamiltonian can be calculated exactly, i.e. without any Taylor approximation. The kinetic terms are easily truncated to two levels with standard methods. We find the following qubit Hamiltonian
\begin{align}
H_{2}^{I} &= -\frac{1}{2}\Omega_{q}(\sigma_{1}^{z} + \sigma_{2}^{z}) - \frac{1}{2}\Omega_{c}\sigma_{c}^{z} \nonumber\\
	&\hspace{11.5pt} - J_{qq}^{z}\sigma_{1}^{z}\sigma_{2}^{z} - J_{qc}^{z}(\sigma_{1}^{z}\sigma_{c}^{z} + \sigma_{2}^{z}\sigma_{c}^{z}) + J_{qqc}^{z}\sigma_{1}^{z}\sigma_{2}^{z}\frac{1 - \sigma_{c}^{z}}{2} \nonumber\\
	&\hspace{11.5pt} + J_{qqc}^{x}\sigma_{1}^{x}\sigma_{2}^{x}\frac{1 - \sigma_{c}^{z}}{2} \label{idealH2}
\end{align}
where the $ \Omega $'s and $ J $'s are positive spin model parameters. Their explicit form can be found in \cref{app:idealrecast}. The circuit has thus resulted in two resonant qubits and a detuned third, coupled via pairwise ZZ-couplings, a three-body ZZC-coupling and a three-body XXC-coupling. C refers to control and to the operator $ (1 - \sigma^{z})/2 $, whose eigenvectors are the ground and excited state of the qubit with eigenvalues $ 0 $ and $ 1 $ respectively. This shows that this type of coupling is turned off and on depending on whether the $ c $-qubit is in its ground or excited state. The ZZC-coupling is small, being sixth order in the fluxes and would thus not have survived a usual expansion to fourth order. Notice how there are no other XX-couplings except the controlled one. The couplings from the capacitances are gone because of the choice of coordinates, and while the Josephson junctions do generally contribute such couplings, they have cancelled because of the symmetry of the circuit parameters. This shows the power of our method. The desired three-body interaction is native in these coordinates and there are no undesired couplings, which we need to deal with. The Z-type couplings merely shift the energy levels of the system, which can be compensated. As we shall see, the XXC-coupling has a strength comparable to the anharmonicities as a consequence of these two quantities arising from the same circuit elements and the same type of terms in the Hamiltonian. More explicitly, it is the same circuit parameters that determine the strength of both these two quantities and they both arise from terms that are fourth order in the node fluxes, meaning they will naturally tend to be similar in size. Hence, the coupling is additionally quite strong, making the time scale very fast, and certain couplings with states outside the qubit subspace will not be suppressed by the anharmonicities. As mentioned we nonetheless consider the degrees of freedom as qubits, though it is important to keep in mind that some of the qubit states are "forbidden", as we shall see, if we want to the system to remain in the qubit subspace. Naming qubits $ 1 $ and $ 2 $ input and output, and the $ c $-qubit control, we see that in a rotating wave approximation the effect of the XXC-coupling is that the input and output will flip each other conditioned on the state of the control qubit. The ZZ- and ZZC-couplings effectively only induce shifts of the qubit energies depending on their states, and since the Hamiltonian is symmetric under the exchange of qubits $ 1 $ and $ 2 $, the states $ \ket{0_{1}1_{2}} $ and $ \ket{1_{1}0_{2}} $ for any state of the $ c $-qubit will still be resonant. These are the states which will be flipped into one another by the XXC-coupling.

\subsection{The implemented gate}\label{sec:ideal:gate}
The qubit transition frequencies are about two orders of magnitude larger than the coupling strengths, as we shall detail later. Therefore interactions that do not preserve the number of excitations will be highly suppressed. In the same way we found exact expressions for spin model parameters in $ H_{2} $, we can find exact expressions for the anharmonicities, $ \alpha_{i} $, which can be seen in \cref{app:idealanharm}. To fourth order they have the following simple relations to each other and the coupling strengths
\begin{align}
\alpha_{1} = \alpha_{2} = \frac{3}{4}\alpha_{c} = \frac{3}{4}J_{qqc}^{x} = 3J_{qc}^{z} = 2J_{qq}^{z}
\end{align}
Hence, the coupling strengths are of same order of magnitude as the anharmonicities, and so interactions with higher energy levels that preserve the number of excitations will not be suppressed, for example $ \ket{1_{1}1_{2}1_{c}} \leftrightarrow \ket{0_{1}2_{2}1_{c}} $. To justify reducing our local degrees of freedom to two-level qubits, we must make sure to not populate the states which couple to outside of the qubit subspace. These "forbidden"\ states are $ \ket{1_{1}1_{2}0_{c}} $ and $ \ket{1_{1}1_{2}1_{c}} $, i.e. those where both input and output qubits are excited. States where at most one of qubits $ 1 $ and $ 2 $ is excited are protected from interaction with higher states by the conservation of excitations and the fact that the $ c $-qubit only changes its excitation in even numbers. This is a consequence of $ \phi_{c} $ only entering in even powers in the Hamiltonian \cref{eq:idealfluxH}. Therefore, if we initialize in a state where there is at most one excitation among qubits $ 1 $ and $ 2 $, we will never populate the $ \ket{1_{1}1_{2}} $ states. Hence, we may consider our system to implement a gate operating on  a six-state Hilbert space consisting of the states $ \ket{0_{1}0_{2}} $, $ \ket{0_{1}1_{2}} $ and $ \ket{1_{1}0_{2}} $ for any state of the $ c $-qubit. The matrix of this gate for the $ c $-qubit in its ground state and in the excited state can be written as
\begin{align}
U_{c=0} = \mat{1 & 0 & 0 \\ 0 & e^{i\beta} & 0 \\ 0 & 0 & e^{i\beta}}, \hspace{5pt} U_{c=1} = \mat{e^{i\gamma} & 0 & 0 \\ 0 & 0 & e^{i\delta} \\ 0 & e^{i\delta} & 0} \label{eq:U}
\end{align}
These two correspond to the interaction being off or on, and likewise we may consider the gate to be closed or open. We have included phases $ \beta $, $ \gamma $ and $ \delta $ (and ignored an irrelevant overall phase), which are a result of different dynamical phases on the different states. These phases depend on the exact energies of the system and therefore on the exact circuit parameters. This means they can be predicted by simulation and even tuned to some extent. Furthermore, working in an appropriately rotating frame, these phases would be zero. From these matrices we see we can implement a CNOT-gate using the states $ \ket{0_{1}1_{2}} $ and $ \ket{1_{1}0_{2}} $ as our logical $ \ket{0} $ and $ \ket{1} $. We can also implement a transistor which transfers quantum information with a gate that is conditioned on our control qubit. This has been discussed recently in Refs. \cite{Marchukov2016,Lloyd2016,Loft2018}. While \cite{Marchukov2016} finds that a minimum of four qubits is required for transistor functionality, the three-body interactions in our system reduces this requirement ot just input, output and gate qubits. The transfer process can also be used to generate entanglement between input and output in the case where the control is in a superposition state.

\begin{figure*}
	\includegraphics[scale=1]{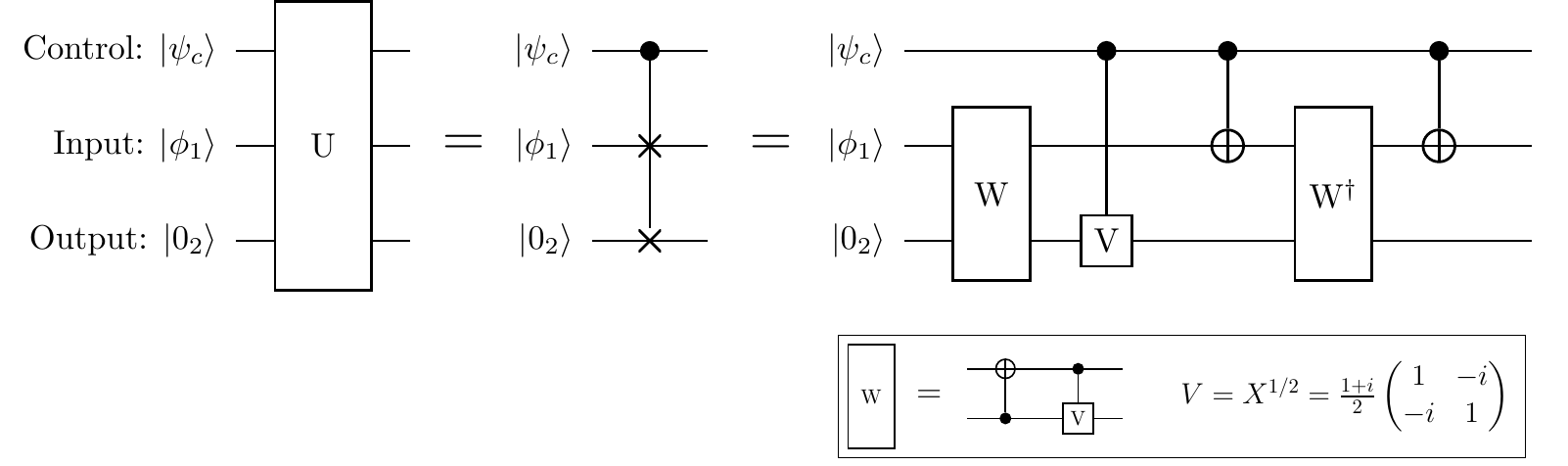}
	\caption{\label{fig:gatedecom} Our gate $ U $ is, ignoring the phases, a Fredkin gate, except we must restrict the output qubit to always be initialized in the ground state $ \ket{0_{2}} $, while the other two qubits can be in any state $ \ket{\phi_{1}\psi_{c}} $. The Fredkin gate is decomposed into five two-qubit gates \cite{Smolin1996}, where the W-gate is the conjunction of a controlled-NOT and a controlled-V, where V is the square root of a Pauli X-gate.}
\end{figure*}

In a frame where the phases vanish, this gate is similar to a Fredkin gate \cite{NielsenChuang2010}, also called CSWAP, except we must restrict the initial state of the output qubit to always be $ \ket{0_{2}} $. We call the resulting gate a restricted Fredkin gate. In Ref. \cite{Smolin1996} they show a decomposition of the Fredkin gate into five two-qubit gate, which has been proven to be the minimal number of gates necessary \cite{Yu2015}. The decomposition of the gate can be seen in \cref{fig:gatedecom}. We have introduced the two-qubit gate W, which is the conjunction of a controlled-NOT and controlled-V, where V satisfies V = X$ ^2 $, as indicated in the figure. This shows the advantage of using a direct implementation of a three-qubit gate. An experimental implementation using no fewer than five gates would likely induce more noise, would have a longer operation time, and would simply require more hardware. Furthermore, the W gate is not a standard two-qubit gate and would require additional work to implement directly. By rethinking quantum circuits and algorithms in terms of many-qubit gates, we can greatly decrease complexity and the implementational issues that follow.  However, it should be noted that it has been shown that some multi-qubit gates can have shorter operation times, when decomposed, than the sum of the operation times of their components \cite{Ashhab2012}. Thus, though it appears that the decomposed Fredkin gate might have an operation time around five times that of a standard two-qubit gate, like the CNOT-gate, it could be significantly less. Nonetheless, it seems intuitive that a direct multi-qubit gate would be superior, and this possibility demands study.

\section{Realistic system}\label{sec:real}
\subsection{External coupling}\label{sec:real:ext}
We now add capacitive couplings to external control and ground. The external control could consist of resonators for readout, flux lines for tuning, and drive lines for preparation \cite{Touzard2019}. Each external coupling adds to the diagonal of the capacitance matrix, and we assume that the total contribution for each node is identical and equal to $ K $. The external couplings are drawn in red in \cref{fig:circuit} as a single capacitor from each node coupling to outside of the circuit. Hence, we assume that these external couplings are identical for each node in the circuit. If they are not identical it will induce interactions between the modes after the coordinate transformation. The interaction strengths will be determined by the discrepancies between the capacitances of the external couplings. However, whether or not they are identical, we can drive each of our modes individually by essentially performing the same transformation of the drives as the coordinates. The general term in the Hamiltonian for some external capacitive coupling will have the form $ -K_{j}D_{j}\dot{\psi}_{j} $, where $ K_{j} $ is the capacitance of the coupling and $ D_{j} $ is the externally controlled variable. This variable could be a voltage, flux or current. Gathering these variables in a vector $ \textbf{D}_{\psi} $, we can write all couplings together as
\begin{align}
H_{ext} = -\textbf{D}_{\psi}^{T}\mathcal{K}_{D}\dot{\bm{\psi}}
\end{align}
where $ \mathcal{K}_{D} $ is a diagonal matrix with $ K_{j} $ as entries. If these are all identical and equal to $ K $, and we write $ \textbf{D}_{\psi} = V\textbf{D}_{\phi} $ for some other drives $ \textbf{D}_{\phi} $, then after the coordinate transformation we have
\begin{align}
H_{ext} = -K\textbf{D}_{\phi}^{T}\dot{\bm{\phi}}
\end{align}
Hence, the drives $ \textbf{D}_{\phi} $ couple individually to the modes $ \bm{\phi} $. Even if the $ K_{j} $ are not identical we can choose $ \textbf{D}_{\psi} $ to achieve the same individual coupling, (see \cref{app:readout}), but we will as mentioned have additional XX-couplings between the modes.

\subsection{Circuit \& Hamiltonian}\label{sec:real:circHamil}
These couplings result in an asymmetry between $ \phi_{1} $ and $ \phi_{2} $, because the entries in the capacitance matrix (after the coordinate transformation) pertaining to these two modes will no longer be identical (see \cref{app:realcirc}). This creates a detuning and the XX-coupling between them no longer cancels completely. While this coupling is small, the detuning is not negligible even for $ K \simeq \SI{1}{\femto\farad} $. Neither can be tuned to zero without removing the external couplings again. To counter this we add an additional branch to our circuit connecting the two ends via an inductor with inductance $ L_{0} $ and a parasitic capacitance $ C_{0} $. To cancel XX-couplings from this new inductor and to ensure that the capacitance network retains the same modes, we must furthermore add an identical inductor to the middle transmon block. This is drawn in blue in \cref{fig:circuit}. The circuit is now similar to that of \cite{Roy2017}, with some non-linear inductors replaced with linear ones.  Notice, that these new inductors do not add any new dynamical degrees of freedom to the physical system but simply alter the parameters of the existing qubits. So while they do add to the complexity of the circuit they should not be considered an overhead or as ancillary, but as an equal and integrated part of the altered circuit. The new capacitances actually add to the asymmetric terms and should therefore be as small as possible. The contributions from the inductors are also asymmetric, and can be used to tune the states $ \ket{0_{1}1_{2}1_{c}} $ and $ \ket{1_{1}0_{2}1_{c}} $ back into resonance, balancing out the asymmetry from the capacitances. The undesired XX-coupling cannot be removed completely, though it can be tuned to be relatively small. We will therefore not be able to completely turn off the XX-coupling between the input and output qubits, in other words we can not completely close the gate. The leakage through the closed gate, when the control qubit in its ground state and the XXC-coupling is thus off, is however quite small as we shall see.

The resulting Hamiltonian is
\begin{align}
H^{R} &= \phantom{+} \frac{\mathcal{C}_{1}^{-1}}{2}p_{1}^2 + \frac{\mathcal{C}_{2}^{-1}}{2}p_{2}^2 + \frac{\mathcal{C}_{c}^{-1}}{2}p_{c}^2 \nonumber\\
&\hspace{11.5pt} - E\cos(2\phi_{1} + 2\phi_{2}) \nonumber\\
&\hspace{11.5pt} - 2E\cos(\sqrt{2}\phi_{1} - \sqrt{2}\phi_{2})\cos(2\phi_{c})\nonumber\\
&\hspace{11.5pt} + 8E_{L_{0}}\left[(2+\sqrt{2})\phi_{1}^2 + (2-\sqrt{2})\phi_{2}^2\right] \label{eq:realfluxH}
\end{align}
where $ \mathcal{C}_{i}^{-1} $ are matrix elements of the inverted capacitance matrix (which are no longer as simple as in \cref{eq:idealfluxH}). Their exact expression can be found in \cref{app:realcirc}. $ E_{L_{0}} = \frac{1}{2L_{0}} $ is the inductive energy of the newly introduced inductors. We can see that there is no interaction pertaining to $ E_{L_{0}} $, as this has cancelled between the two inductors as intended. However, when we truncate the system to three qubits, there is still an uncontrolled XX-coupling between input and output, because of the introduced asymmetry. The qubit Hamiltonian is
\begin{align}
H_{2}^{R} &= -\frac{1}{2}\Omega_{1}\sigma_{1}^{z} - \frac{1}{2}\Omega_{2}\sigma_{2}^{z} - \frac{1}{2}\Omega_{c}\sigma_{c}^{z} \nonumber\\
	&\hspace{11.5pt} - J_{12}^{z}\sigma_{1}^{z}\sigma_{2}^{z} - J_{1c}^{z}\sigma_{1}^{z}\sigma_{c}^{z} - J_{2c}^{z}\sigma_{2}^{z}\sigma_{c}^{z} + J_{12c}^{z}\sigma_{1}^{z}\sigma_{2}^{z}\frac{1 - \sigma_{c}^{z}}{2} \nonumber\\
	&\hspace{11.5pt} - J_{12}^{x}\sigma_{1}^{x}\sigma_{2}^{x} + J_{12c}^{x}\sigma_{1}^{x}\sigma_{2}^{x}\frac{1 - \sigma_{c}^{z}}{2} \label{eq:realH2}
\end{align}
which is the same as with the idealized circuit except for the additional XX-coupling and the fact that the explicit symmetry between qubits $ 1 $ and $ 2 $ is broken. The spin model parameters are calculated analytically in \cref{app:realcirc}. Hence, we expect the same dynamics as before, except the closed case will not be completely closed. Furthermore, we must tune the circuit parameters to make $ \ket{0_{1}1_{2}1_{c}} $ and $ \ket{1_{1}0_{2}1_{c}} $ resonant, compensating partially for non-identical qubit frequencies and for the Z-type couplings which shift the energy levels. 

\section{Simulation}\label{sec:sim}
Let us now move on to numerical calculations and simulations of this circuit. We first show an example of the most important spin model parameters, showing how the couplings are comparable and even larger than the anharmonicities, resulting in extremely fast dynamics. Afterwards we show how the population of the qubit states evolve depending on the initial state, which show that states with zero or one excitation are essentially stationary, $ \ket{0_{1}1_{2}1_{c}} $ and $ \ket{1_{1}0_{2}1_{c}} $ exchange their populations as we want, and $ \ket{1_{1}1_{2}0_{c}} $ and $ \ket{1_{1}1_{2}1_{c}} $ quickly evolve out of the qubit subspace. Finally, we use the average fidelity \cite{Nielsen2002} as a measure of the quality of our implementation of the gate described in \cref{eq:U}. With the altered circuit, we can however not expect the phases acquired by $ \ket{0_{1}1_{2}0_{c}} $ and $ \ket{1_{1}0_{2}0_{c}} $ to be identical. Therefore the two $ \beta $'s in \cref{eq:U} are distinguished from each other by calling them $ \beta_{1} $ and $ \beta_{2} $ respectively. Simulations include noise by using the Lindblad master equation 
\begin{align}
\frac{d}{dt}\rho = -i[H,\rho] + \sum_{k}\gamma_{k}\left(L_{k}\rho L_{k} - \frac{1}{2}\left\{L_{k}^2,\rho\right\}\right)
\end{align}
where $ \rho $ is the density matrix, $ L_{k} $ are Lindblad operators representing  different noise channels with rates $ \gamma_{k} $, and the sum runs over all Lindblad operators. The Lindblad operators are $ \op{\psi}{\psi} $ for each state $ \ket{\psi} $ in the Hilbert space inducing dephasing, and annihilation operators $ b_{i} $ for each anharmonic oscillator $ \phi_{i} $ inducing photon loss.  The decoherence rates are set to the same value for both types of noise and for all states and qubits, $ \gamma_{k} = \SI{0.05}{\mega\hertz} $, giving them a lifetime of $ \SI{20}{\micro\second} $, which is well within state-of-art achievements \cite{Wang2019,Touzard2019}.  We simulate four levels for each non-linear oscillator to include the (suppressed) interactions with higher levels. In addition to causing two of the qubit states to evolve out of the qubit subspace, these interactions cause effective shifts of the energies of the qubit states. This is similar to the effects of higher levels considered in Ref. \cite{Ashhab2008}. These shifts can essentially be considered as additional Z-type couplings and can be compensated for as mentioned previously. All simulations make use of the \textsc{Python} toolbox \textsc{QuTiP} \cite{QuTip2013}.

\subsection{Spin model parameters comparison}\label{sec:sim:smp}
\begin{figure}
	\centering
	\includegraphics[width=\columnwidth]{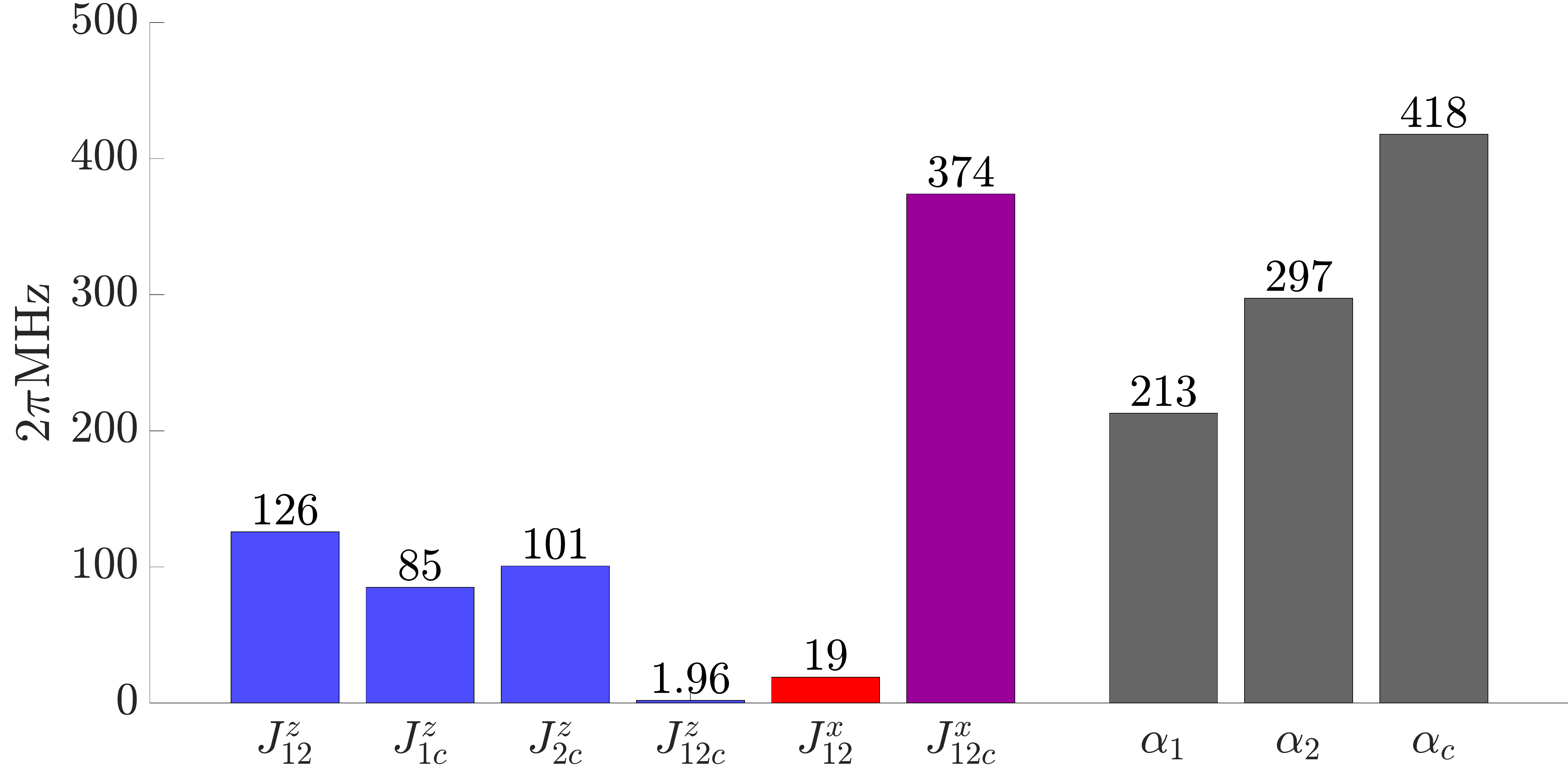}
	\caption{\label{fig:smp} The most important spin model parameters. The ZZ- and ZZC-coupling strength in blue, the XX-coupling strength in red, the XXC-coupling strength in purple and the anharmonicities in grey. The qubit transition frequencies $ \Omega_{i}/2\pi $ are all close to $ \SI{36}{\giga\hertz} $.}
\end{figure}
In our numerical search for circuit parameters, we fix $ C_{0} $ and $ K $ at $ \SI{1}{\femto\farad} $, and vary $ C $, $ E $, $ L_{0} $ to satisfy the following conditions: the states $ \ket{0_{1}1_{2}1_{c}} $ and $ \ket{1_{1}0_{2}1_{c}} $ must be resonant; the anharmonicities must be greater than $ 200\times\si{2\pi\mega\hertz} $; the ratio between the kinetic and harmonic potential terms for each qubit must be small to ensure that we are in the transmon regime (see \cref{app:idealrecast}); the ratio $ J_{12}^{x}/J_{12c}^{x} $ must be small. The detuning between $ \ket{0_{1}1_{2}1_{c}} $ and $ \ket{1_{1}0_{2}1_{c}} $ is at first calculated from the analytical expressions for their energies, and for higher precision it is then calculated from the infidelity of the process $ \ket{0_{1}1_{2}1_{c}} \leftrightarrow \ket{1_{1}0_{2}1_{c}} $. The analytical expressions do not completely capture the detuning because of higher order energy shifts from the strong ZZ-couplings  and from interactions with higher levels as mentioned. An example of the spin model parameters can be seen in \cref{fig:smp}. In this simulation we have $ E/2\pi = \SI{200}{\giga\hertz} $, $ C = \SI{21.67}{\femto\farad} $ and $ L_{0} = \SI{12.67}{\nano\henry} $. We require such a large value of $ E/2\pi $ in order to make the contributions from the linear inductors relatively small, as these must balance out the asymmetric contributions from the capacitors, but not become dominant themselves. The size of the asymmetric inductive contribution is given by the ratio $ 2E_{L_{0}}/E $, while $ (2C_{0} + K)/2C $ determines the size of the asymmetric capacitive contribution (see \cref{app:realcirc}). For the parameters given here we have $ E_{L_{0}}/E = 0.065 $ and $ (2C_{0} + K)/2C = 0.069 $. We can see that as expected $ J_{12c}^{z} $, which is sixth order in the fluxes, is much smaller than the others. However, both the ZZ-couplings and in particular $ J_{12c}^{x} $ are comparable to the anharmonicities. This large value of $ J_{12c}^{x} $ gives us very fast dynamics, but clearly the anharmonicities are not large enough to suppress interactions that have coupling strengths like these. Lastly, $ J_{12}^{x} $ is small, but unfortunately not negligible, which will reduce the average fidelity of the gate, as the closed case, where the control qubit is in the state $ \ket{0_{c}} $, will not be entirely closed. In this example the three qubit transition frequencies $ \Omega_{i}/2\pi $ are all close to $ \SI{36}{\giga\hertz} $, so interactions that do not preserve the number of excitations are heavily suppressed as expected. As mentioned, the ZZ- and XXC-couplings are of the same order of magnitude as the anharmonicities, because they arise from the same terms of the original flux Hamiltonian \cref{eq:realfluxH}, i.e. fourth and higher order terms from the cosines.  As we must tune the anharmonicities to be large enough so that we can experimentally address the individual qubits for preparation and so on, the couplings also become large.

\subsection{Population evolution}\label{sec:sim:pop}
\begin{figure*}
	\centering
	\includegraphics[width=\textwidth]{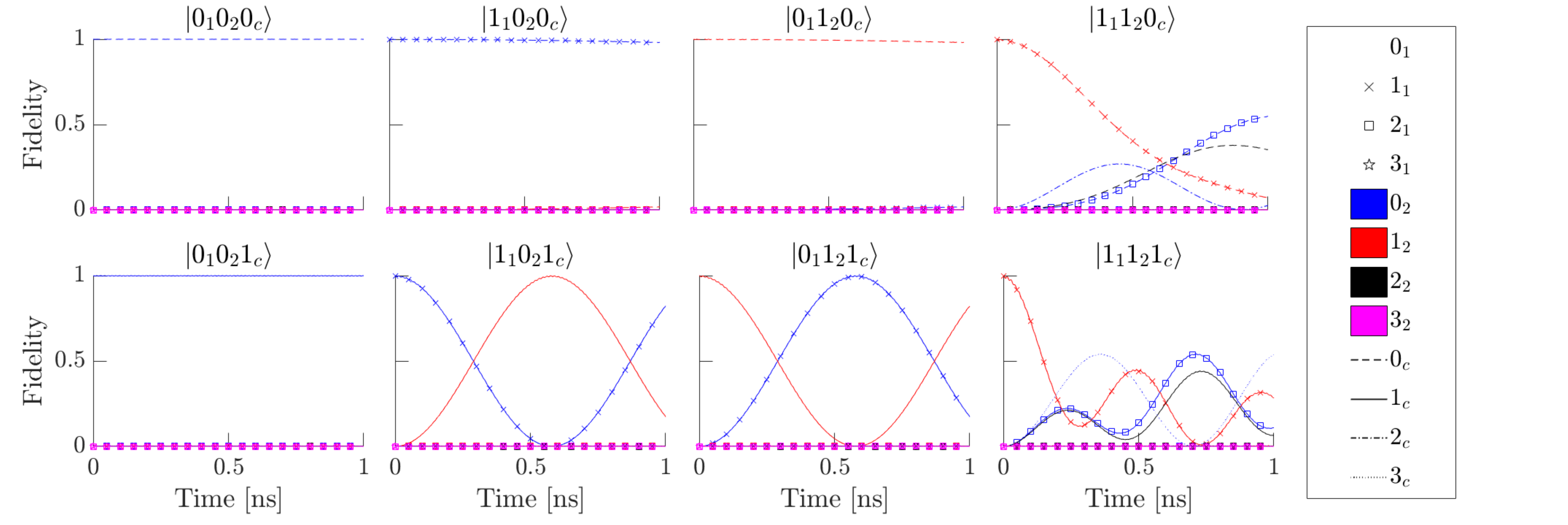}
	\caption{\label{fig:all8} The state populations of the non-linear oscillators simulated with four levels each. The titles indicate the initial state, with all closed states in the upper row and all open in the lower. The marker, colour and line type of a curve indicate which state it pertains to, as described in the legend on the right.  For example, crosses ($ 1_{1} $) on a blue ($ 0_{2} $) solid ($ 1_{c} $) curve represent the population of $ \ket{1_{1}0_{2}1_{c}} $, while for example squares ($ 2_{1} $) on a magenta ($ 3_{2} $) dashed ($ 0_{c} $) curve represent $ \ket{2_{1}3_{2}0_{c}} $. We see that the states with zero or one excitation are nearly stationary, while $ \ket{1_{1}0_{2}1_{c}} $ and $ \ket{0_{1}1_{2}1_{c}} $ exchange their populations. Finally the states with both input and output excited quickly evolve out of the qubit subspace. }
\end{figure*}
We simulate the dynamics of the system with the parameters presented in the previous section. In \cref{fig:all8} can be seen the populations of all states as we initialise the system in one of the eight qubit states. Clearly $ \ket{0_{1}0_{2}0_{c}} $, $ \ket{1_{1}0_{2}0_{c}} $, $ \ket{0_{1}1_{2}0_{c}} $ and $ \ket{0_{1}0_{2}1_{c}} $ are nearly stationary, i.e. their populations do not change, though they may acquire a phase. The states $ \ket{0_{1}1_{2}1_{c}} $ and $ \ket{1_{1}0_{2}1_{c}} $ exchange their populations, as the XXC-coupling is on, corresponding to the input and output qubits flipping each other. The very large value of $ J_{12c}^{x} $ has resulted in dynamics on the nanosecond scale. Lastly, $ \ket{1_{1}1_{2}0_{c}} $ and $ \ket{1_{1}1_{2}1_{c}} $ quickly evolve out of the qubit space and excite a mess of higher states.  Simulating the system with the same circuit parameters but including six levels for each anharmonic oscillator makes no significant difference, indicating that we have captured the effects of the higher levels by taking just four levels into account. Additionally, it can be seen that we have properly compensated for the effective energy shifts of the energy levels caused by higher levels and the strong couplings in the system. Finally, we find in our simulations that the qubit subspace loses $ 0.6\% $ or less of its population due to leakage to the higher levels, except of course when initializing in the states $ \ket{1_{1}1_{2}0_{c}} $ or $ \ket{1_{1}1_{2}1_{c}} $, in which case most of the population goes to the higher levels after some time.

\subsection{Average fidelity}\label{sec:sim:fid} 
\begin{figure}
	\centering
	\includegraphics[width=\columnwidth]{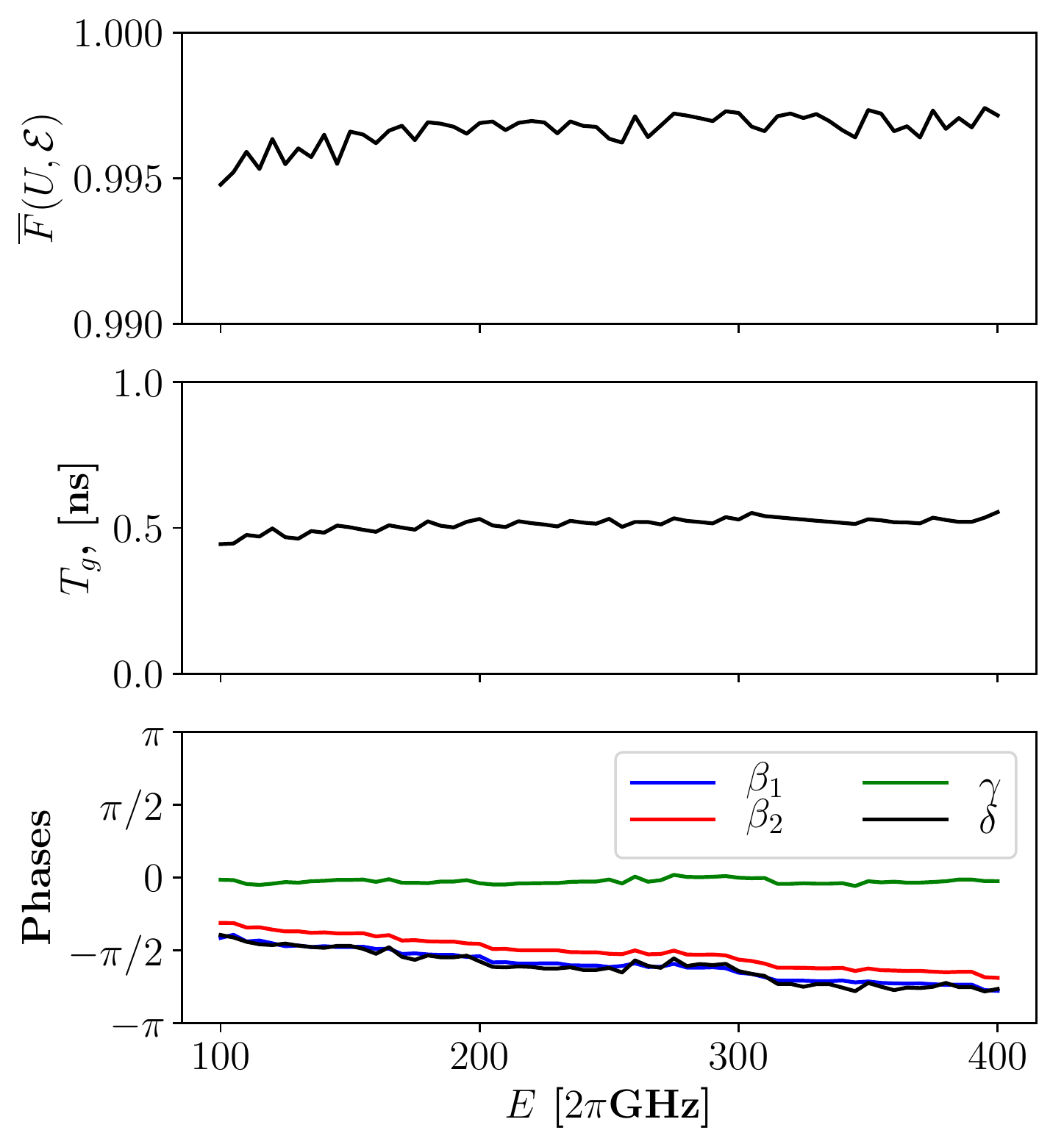}
	\caption{\label{fig:iter} Plots of \textbf{(a)} the average fidelity for our implementation of the gate in \cref{eq:U}, \textbf{(b)} the gate operation time, and \textbf{(c)} the phases of the gate as functions of the Josephson energy $ E $, with $ C $ and $ L_{0} $ determined by maximising the fidelity of $ \ket{0_{1}1_{2}1_{c}} \leftrightarrow \ket{1_{1}0_{2}1_{c}} $.}
\end{figure}

The average fidelity can be calculated from \cite{Nielsen2002}
\begin{align}
\overline{F}(U,\mathcal{E}) = \frac{\sum_{j}\tr{UU_{j}^{\dagger}U^{\dagger}\mathcal{E}(U_{J})} + d^2}{d^2(d + 1)}
\end{align}
where $ U $ is the target gate, and $ \mathcal{E} $ is the quantum map used to implement it. The $ U_{j} $ are a unitary basis for operators on the relevant space of states, which has dimension $ d $. In our case the target gate is the one described in \cref{eq:U} (with $ \beta $ replaced by $ \beta_{1} $ and $ \beta_{2} $), and $ \mathcal{E}(\rho(0)) = \rho(t) $ is the time evolution of our system. The relevant space of states is spanned by the six qubit states that do not evolve out of the qubit subspace. Thus for any set of circuit parameters and phases $ \beta_{1} $, $ \beta_{2} $, $ \gamma $ and $ \delta $, we can calculate the average fidelity as function of time. We choose the phases to maximise the fidelity and call the time of maximum fidelity the gate operation time $ T_{g} $. This time is expected to be approximately $ \frac{\pi}{2(J_{12c}^{x} + J_{12}^{x})} $, as we are essentially considering the oscillation between two resonant energy levels interacting with a coupling strength $ J_{12c}^{x} + J_{12}^{x} $ (see \cref{app:idealrecast}). However, higher order interactions make $ T_{g} $ slightly smaller than this, and so we determine it numerically. We do this for a range of circuit parameters, choosing values for $ E/2\pi $ in $ [100,400]\SI{}{\giga\hertz} $, and tuning $ C $ and $ L_{0} $ to satisfy the demands described above, in particular maximising the fidelity of the process $ \ket{0_{1}1_{2}1_{c}} \leftrightarrow \ket{1_{1}0_{2}1_{c}} $. The results can be seen in \cref{fig:iter}. These are simulated without noise. The fidelities are above $ 99.5\% $ and vary only over a short interval, increasing slightly with the value of $ E $. There are small, noisy variations in the curve, attributed to numerical imprecision and randomness in the optimization routine. In particular the calculations are affected by there being many local minima close together in the space of phases, and in simulating dynamics we have only a finite resolution in time. The gate times are well below $ \SI{1}{\nano\second} $ and increase slightly with $ E $. Finally the angles also vary slightly with $ E $.

\section{Discussion}\label{sec:discu}
The fundamental reason that two of the qubit states are lost, due to interactions with higher levels, is that we have an odd interaction induced by the same circuit components that create the anharmonicities. By odd interaction we mean an interaction stemming from product terms of odd powers of different flux coordinates, like $ \phi_{1}\phi_{2} $ or $ \phi_{1}^3\phi_{2} $, which exchange odd numbers of excitations. Such interactions with higher levels must be suppressed by the anharmonicities. Since these interactions come from the same circuit elements as the anharmonicities (and both are fourth order in the flux coordinates), they inherently are comparable in strength, and so the anharmonicities can not suppress them. This is in opposition to even interactions with higher levels, i.e. interactions stemming from terms like $ \phi_{1}^2\phi_{2}^2 $, that exchange even numbers of excitations, and which are suppressed by the approximate conservation of the number of excitations. It would require a qubit to have for example two excitations in order for another qubit to be twice excited, but we always initialize in states with only zero or one excitation. We could therefore employ the above method of a symmetric circuit viewed in the basis of electrical modes more successfully if the odd interactions came from different elements than the anharmonicities, such that they could possibly be tuned independently, or if we only had even interactions, as is done in Refs. \cite{Roy2017,Roy2018}. 

The gate we have proposed works on the scale of a single nanosecond, which is very fast compared to other present two- or multi-qubit gates. This may pose a challenge for current implementation, but we expect our implementation to be relevant in the future as other operation times are expected to decrease. We note in passing that the one-qubit operation times can be $ \SI{10}{\nano\second} $ or shorter \cite{Chow2010,Deng2015} and our gate times are comparable. As mentioned, recently a two-qubit gate with operation time of less than a nanosecond was experimentally achieved in the context of electron spin qubits in atoms \cite{He2019}. Our proposal predicts comparable results for superconducting circuits, though for a direct three-qubit gate working in a limited part of the qubit subspace. Furthermore, joining several fast gates together in a network might result in a useful component that takes advantage of the fast operations. Alternatively to a three-qubit gate, direct three-body interactions are useful for quantum simulation of lattice gauge theories. For example an XXX-coupling is needed for the hopping of particles between lattice sites mediated by a U(1) gauge field living on the link between sites \cite{Hauke2013}, and the matter-gauge interaction in $ \mathbb{Z}_{2} $ gauge theories is an XXZ-coupling \cite{Prosko2017}. Our approach could be used to derive circuits which implements such interactions directly. Indeed, we found an XXZ-coupling in the presented circuit. Likewise, three-body interactions could be used to implement three-local constraints in the context of quantum annealing schemes. Here it is most often even interactions, like a ZZZ-coupling, which are used. Such an interaction could certainly also be found with our approach, indeed we did in fact find it in the above circuit and wrote it as a ZZC-coupling. It would, however, necessarily be small because it would be sixth order in the fluxes, just as we found $ J_{12c}^{z} $ to be much smaller than the other coupling strengths. If one could figure out a way for other couplings to cancel or become irrelevant, it might still be a good approach and faster than indirect implementations.

Alternatively to the above work, one could interpret the system as a qutrit, consisting of the states $ \ket{0_{1}0_{2}} $, $ \ket{0_{1}1_{2}} $ and $ \ket{1_{1}0_{2}} $, interacting with the $ c $-qubit. The energy levels $ \ket{0_{1}1_{2}} $ and $ \ket{1_{1}0_{2}} $ of the qutrit will be degenerate and, depending on the state of the $ c $-qubit, will either be stationary or oscillate between each other. Or one could diagonalize the two higher levels of the qutrit with the states $ \frac{1}{\sqrt{2}}\left(\ket{0_{1}1_{2}} \pm \ket{1_{1}0_{2}}\right) $, which will again either be degenerate or have an energy splitting of $ 2J_{12c}^{x} $, depending on whether the XXC-coupling is off or on. Proposals have been made on how to utilize the third level of a qutrit to reduce the number of elementary gates necessary to implement concrete quantum computations \cite{Ralph2007,Bishop2008,Miranowicz2014,Leonski2001,Liu2005,Baekkegaard2019,Fedorov2011,Lanyon2008,Luo2014_Wang,Li2013,Luo2014_Yang}.

However, it is also possible to employ a set up, which actively chooses to only make use of exactly the states we have available in our system. In Refs. \cite{Lloyd2016,Ciani2019} two physical qubits represent a site in a lattice, which can host a mobile logical qubit. The state $ \ket{00} $ represents a vacant site, while the states $ \ket{01} $ and $ \ket{10} $ represent the two internal states of a present logical qubit. In this way the logical qubits are able to move around the lattice while carrying quantum information.

\subsection{Scalability}\label{sec:scal}
\begin{figure}
	\centering
	\includegraphics[width=0.9\columnwidth]{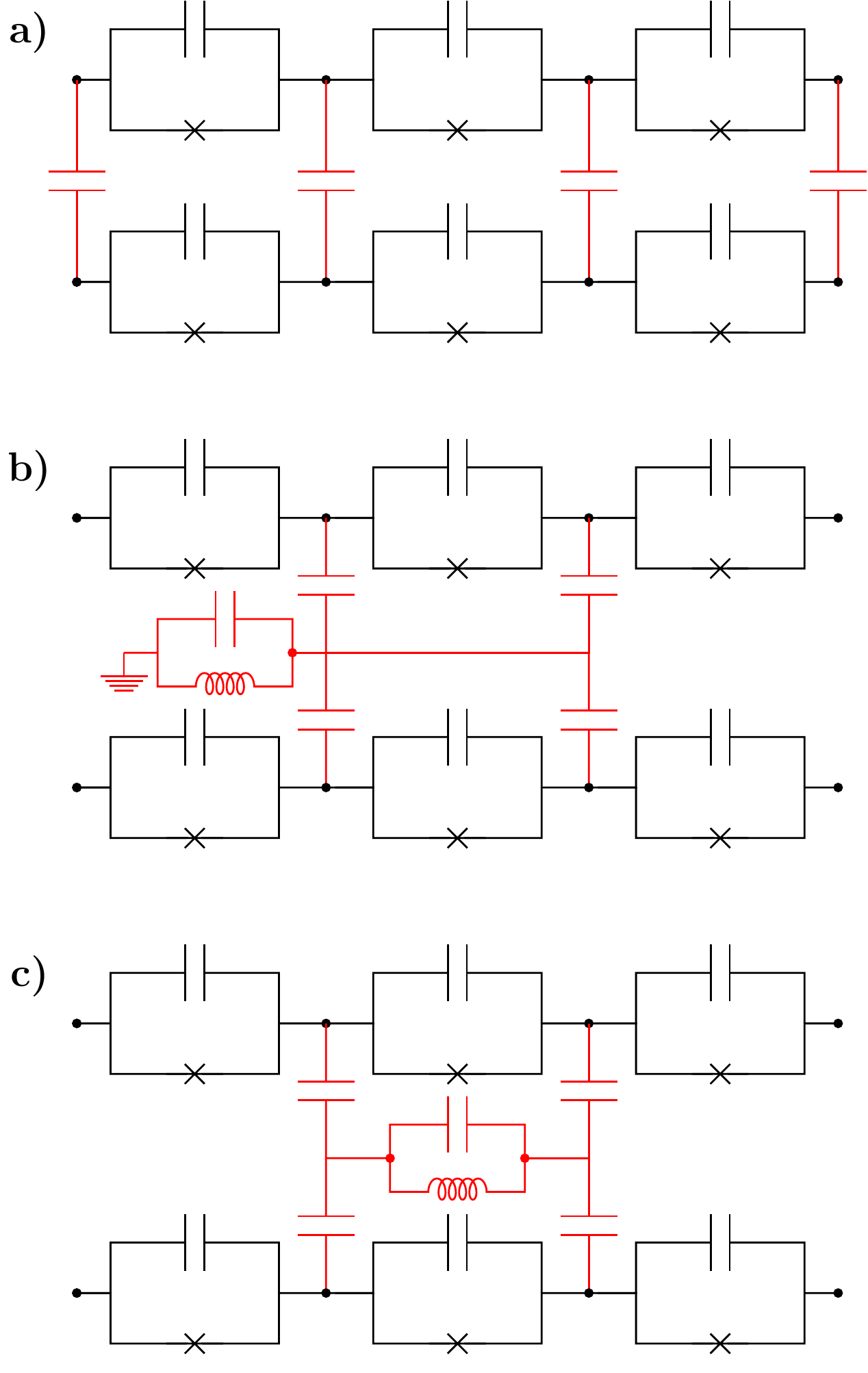}
	\caption{\label{fig:scal}  Three different approaches for scaling of the circuit (in black), always using identical capacitative couplings to either a copy of the circuit or a mediating resonator (in red). \textbf{a)} Using capacitors between every node and its counterpart in the copied circuit couples each mode in the first circuit to its counterpart in the copied circuit. \textbf{b)} Connecting the middle two nodes of the circuit to a mediating resonator couples the modes $ \phi_{CM} $ and $ \phi_{c} $ to the resonator. \textbf{c)} Connecting the middle two nodes of the circuit to the ends of a mediating resonator couples the modes $ \phi_{1} $ and $ \phi_{2} $ to the resonator.}
\end{figure}
Finally, we consider scaling of the system by looking at the idealized circuit, and how it might be coupled to copies of itself. The challenge that needs to be addressed in regards to scaling is the fact that the flux coordinates $ \bm{\phi} $ are all a linear combination of all of the original node fluxes $ \bm{\psi} $. Hence, individual coupling any of these degrees of freedom, and the corresponding qubits, to external qubits will require coupling to all four nodes in our circuit. However, connecting pairs of nodes to external circuits can be used to couple pairs of the modes to the external degrees of freedom. 

In \cref{fig:scal} can be seen three approaches to scaling the idealized circuit by connecting it to a copy of itself, either directly or through a mediating resonator. The first approach is to connect each node of the circuit with identical capacitors to the corresponding nodes of the copy. This results in XX-couplings between each of the modes and their corresponding partners in the copy. The second approach connects the two middles nodes with identical capacitors to a resonator. This results in the modes $ \phi_{CM} $ and $ \phi_{c} $ coupling to the resonator. Likewise, the third approach connects the two middles nodes with identical capacitors to the two ends of a resonator. This results in the modes $ \phi_{1} $ and $ \phi_{2} $ coupling to the resonator. If might then be possible to tune the resonator and the couplings between it and the modes to mediate some interesting coupling between the main circuits.

Implementing scaling via a resonator, however, comes with the complication that the two middle nodes of the main circuits now get capacitative contributions which the end nodes do not. In order to retain the proper electrical modes of the main circuits, some capacitors would need to be added to the end nodes. Alternatively, one could also implement scaling by connecting the end nodes via resonators in the same way as the above examples which use the middle nodes. In the context of scaling the main circuit to a long chain of copies connected through mediating resonators, one might utilize an alternating scheme where the mediating resonators are alternatingly coupled to the middle nodes and to the end nodes, in order to ensure identical capacitative contributions to all nodes of the main circuits.

\section{Conclusion}\label{sec:conc}
In this article we have seen how a symmetric superconducting circuit of three transmon blocks in series can implement a direct three-body interaction between its electrical modes. The modes are non-linear oscillators and can be considered as qubits. Specifically, we found a strong controlled XX-coupling between two of these qubits conditioned on the state of the third. This interaction can be used to implement a CNOT-gate or a transistor among quantum degrees of freedom  using what we call a restricted Fredkin gate. However, as this coupling is induced by the same circuit elements as the anharmonicities of the non-linear oscillators, the coupling strength is comparable to the anharmonicities. This means that some interactions with higher levels are not suppressed. Specifically, the two states where both the input and output qubits are excited will quickly evolve out of the qubit subspace. Furthermore, external couplings to control and ground induce an asymmetry in the system, which we compensate for by altering the circuit with two linear inductors. The XX-coupling is no longer completely controlled after this, but we still find average fidelities for our implementation of the restricted Fredkin gate above $ 99.5\% $ with operation times below $ \SI{1}{\nano\second} $. The extremely fast time scale is caused by the large coupling strengths.

In regard to scaling of the circuit the challenge that needs to be addressed is the fact that the flux coordinates $ \bm{\phi} $ are all a linear combination of all of the original node fluxes $ \bm{\psi} $. Thus, individual coupling of the qubits to external qubits require coupling to all four nodes in the circuit. This means that cross-talk may be involved and that addressing individual logical qubits in the system requires a careful tuning of couplings to different access points. We have given some examples of how scaling can be approached, either coupling the circuit directly or through a mediating resonator.

\section*{Acknowledgements}
The authors would like to thank N. J. S. Loft, S. E. Rasmussen, L. B. Kristensen and T. Bækkegaard for discussions and idea-development pertaining to this work. The authors acknowledge support from the Independent Research Fund Denmark, the Carlsberg Foundation, and AUFF through the Jens Chr. Skou fellowship program.

\newpage
\onecolumngrid 
\appendix

\section*{Appendix}\label{app}
\subsection{Circuit, Lagrangian and Hamiltonian} \label{app:idealcirc}
The idealized circuit we work with can be seen in black in \cref{fig:circuit} of the main text. To each node we associate a flux coordinate $ \psi_{j} $ for $ j = 1,2,3,4 $ and order these in a vector $ \bm{\psi} = (\psi_{1},\psi_{2},\psi_{3},\psi_{4})^{T} $. If we write up the capacitance matrix for the circuit
\begin{align}
\mathcal{K} = \mat{C & -C & 0 & 0 \\ -C & 2C & -C & 0 \\ 0 & -C & 2C & -C \\ 0 & 0 & -C & C}
\end{align}
we can write the kinetic terms of the Lagrangian as
\begin{align}
\mathcal{L}_{\textup{kin}} = \frac{1}{2}\dot{\bm{\psi}}^{T}\mathcal{C}\dot{\bm{\psi}}
\end{align}
The potential part of the Lagrangian from the Josephson junctions is
\begin{align}
\mathcal{L}_{\textup{pot}} = E\left(\cos(\psi_{1} - \psi_{2}) + \cos(\psi_{2} - \psi_{3}) + \cos(\psi_{3} - \psi_{4})\right)
\end{align}
We now wish to change coordinates to the eigenmodes of the capacitance network, in order to remove interactions from the capacitances. This means we change to the coordinates where $ \mathcal{K} $ becomes diagonal. The matrix used for this change of basis has the eigenvectors of $ \mathcal{K} $ as its columns
\begin{align}
V = \mat{1 & 1+\sqrt{2} & 1-\sqrt{2} & 1 \\ 1 & 1 & 1 & -1 \\ 1 & -1 & -1 & -1 \\ 1 & -1-\sqrt{2} & -1+\sqrt{2} & 1}
\end{align}
Notice that the eigenvectors are not normalized, their norms are (from left to right) $ 4 $, $ 4 $, $ 4(2+\sqrt{2}) $ and $ 4(2-\sqrt{2}) $. The eigenvalues of these are $ 0 $, $ 2C $, $ (2-\sqrt{2})C $ and $ (2+\sqrt{2})C $. If we name the new coordinates $ \phi_{CM} $, $ \phi_{1} $, $ \phi_{2} $ and $ \phi_{c} $, and gather them in a vector $ \bm{\phi} $, we can relate $ \bm{\psi} $ and $ \bm{\phi} $ via
\begin{align}
\bm{\psi} = V\bm{\phi}
\end{align}
Then the kinetic part of the Lagrangian becomes
\begin{align}
\mathcal{L}_{\textup{kin}} = \frac{1}{2}\dot{\bm{\phi}}^{T}V^{T}\mathcal{K}V\dot{\bm{\phi}} = \frac{1}{2}\dot{\bm{\phi}}^{T}\mathcal{C}\dot{\bm{\phi}}
\end{align}
where 
\begin{align}
\mathcal{C} = \mat{0 & 0 & 0 & 0 \\ 0 & 8C & 0 & 0 \\ 0 & 0 & 8C & 0 \\ 0 & 0 & 0 & 8C}
\end{align}
is the transformed capacitance matrix. Notice how the norms of the eigenvectors match with their corresponding eigenvalues in such a way that the non-zero entries are identical. The potential part of the Lagrangian becomes
\begin{align}
\begin{split}
\mathcal{L}_{\textup{pot}} &= E\left(\cos(\sqrt{2}\phi_{1} - \sqrt{2}\phi_{2} + 2\phi_{c}) + \cos(2\phi_{1} + 2\phi_{2}) + \cos(\sqrt{2}\phi_{1} - \sqrt{2}\phi_{2} - 2\phi_{c})\right)  \\
&= E\left(2\cos(\sqrt{2}\phi_{1} - \sqrt{2}\phi_{2})\cos(2\phi_{c}) + \cos(2\phi_{1} + 2\phi_{2})\right)
\end{split}
\end{align}
We see that $ \phi_{CM} $ does not enter at all in the Lagrangian. This mode is exactly like a center of mass-mode, or a constraint variable. It corresponds to a free particle that does not affect the rest of the system. As it does not enter at all, we simply ignore it, but without changing our notation (so for example now $ \bm{\phi} = (\phi_{1},\phi_{2},\phi_{c})^{T} $ and $ \mathcal{C} $ has lost its first row and column). We can now define momenta conjugate to the surviving coordinates
\begin{align}
\textbf{p} = \frac{\p\mathcal{L}_{\textup{kin}}}{\p\dot{\bm{\phi}}} = \mathcal{C}\dot{\bm{\phi}}
\end{align}
With these we can write the kinetic Lagrangian in terms of the momenta 
\begin{align}
\mathcal{L}_{\textup{kin}} = \frac{1}{2}\textbf{p}^{T}(\mathcal{C}^{-1})^{T}\mathcal{C}\mathcal{C}^{-1}\textbf{p} = \frac{1}{2}\textbf{p}^{T}\mathcal{C}^{-1}\textbf{p} 
\end{align}
where we have used that the inverse of $ \mathcal{C} $ is symmetric, because $ \mathcal{C} $ is itself symmetric. As $ \mathcal{C} $ is diagonal, with all diagonal entries non-zero, it is trivially invertible. Incidentally $ \mathcal{C} $ was not invertible before we removed $ \phi_{CM} $. This is because the basis of node fluxes is overdetermined by exactly $ 1 $, i.e. there is one more coordinate than there are degrees of freedom. This is usually dealt with by naming one of the nodes the ground node, and setting the corresponding flux to zero. We have essentially done the same with $ \phi_{CM} $, except it does not simply correspond to one of the nodes in the circuit, but in fact relates to all. As ground also has a more physical meaning, and it is difficult to avoid coupling to it, we will consider a small capacitive coupling to a ground node later on in the more realistic instance of this circuit. As mentioned in the main text this coupling introduces a kind of asymmetry in our system, which we must compensate for.

Finally, with the Lagrangian fully transformed and written in terms of momenta, we can write up the Hamiltonian as 
\begin{align}
\begin{split}
H &= \mathcal{L}_{\textup{kin}} - \mathcal{L}_{\textup{pot}}  \\
&= \frac{E_{C}}{2}\left(p_{1}^2 + p_{2}^2 + p_{c}^2\right) - E\left(2\cos(\sqrt{2}\phi_{1} - \sqrt{2}\phi_{2})\cos(2\phi_{c}) + \cos(2\phi_{1} + 2\phi_{2})\right)
\end{split}
\end{align}
where we have introduced the charging energy $ E_{C} = \frac{1}{8C} $ (in units $ \hbar = 2e = 1 $). Let us introduce $ H_{\textup{kin}} = \mathcal{L} $ and $ H_{\textup{pot}} = -\mathcal{L}_{\textup{pot}} $, as we do not need to refer to the Lagrangian from this point on. If we consider the $ \phi_{i} $ to be position-like coordinates, our system can be considered as three particles interacting through the cosine potentials. In that case we see that the diagonal elements of the capacitance matrix correspond to masses. This can be a helpful way to consider the system, and we will sometimes use language appropriate to it.

\subsection{Recasting and truncating}\label{app:idealrecast}
We now move on to recast our Hamiltonian in terms of harmonic oscillators. The Hamiltonian of a harmonic oscillator is
\begin{align}
H_{HO} = K_{p}p^2 + K_{\phi}\phi^2
\end{align}
where $ \phi $ is a position-like coordinate (in our case a flux) and $ p $ is its conjugate momentum. We promote $ \phi $ and $ p $ to operators and give them the canonical commutation relation $ [\phi,p] = i $. We then define $ r = (K_{p}/K_{\phi})^{1/4} $ and introduce creation/annihilation operators $ a^{\dagger} $/$ a $ related to $ \phi $ and $ p $ via
\begin{align}
\phi  = \frac{r}{\sqrt{2}}(a^{\dagger} + a), \qquad p = i\frac{1}{\sqrt{2}r}(a^{\dagger} - a)
\end{align}
A brief calculation turns the Hamiltonian into the familiar form
\begin{align}
H = 2K_{phi}r^2(2a^{\dagger}a + 1)
\end{align}
Let us do this for our Hamiltonian and all three flux coordinates. We promote all $ \phi_{i} $ and $ p_{i} $ to operators and introduce for each of them creation/annihilation operators $ a_{i}^{\dagger} $/$ a_{i} $. As the Hamiltonian does not explicitly contain any $ \phi_{i}^2 $-terms, we look at the second order terms from the expansion of the cosines in order to define $ r_{i} $. The second order terms alone are
\begin{align}
E\left(2\left(\phi_{1}^2 + \phi_{2}^2 - \frac{1}{2}\phi_{1}\phi_{2}\right) + 2\left(\phi_{1}^2 + \phi_{2}^2 + \frac{1}{2}\phi_{1}\phi_{2}\right) + 8\phi_{c}^2\right) = 4E\left(\phi_{1}^2 + \phi_{2}^2 + \phi_{c}^2\right)
\end{align}
Notice, how the interaction terms cancel neatly. We will later see that all but the desired XXC-coupling and the ZZ-couplings cancel. With this we can define $ r_{i} $, which turn out to be identical in this very neat and symmetric case
\begin{align}
r_{i} = r = \left(\frac{E_{C}}{8E}\right)^{1/4}
\end{align}
With this we can relate $ \phi_{i} $ and $ p_{i} $ to their respective creation/annihilation operators as in the case of a single harmonic oscillator. Terms of the form $ a^2 $ and $ (a^{\dagger})^2 $ will cancel as in the case of a harmonic oscillator, but the fact that we are not actually working with harmonic oscillators is revealed by the presence of terms like $ a^4 $ and $ (a^{\dagger})^4 $, which induce "rotations"\ between different energy levels, without any actual interaction taking place. This reveals that the harmonic oscillator basis is not the natural one for this system, but these rotations will be highly suppressed by a rotating wave-type argument. As we wish to be in the transmon regime, where the fluxes vary only slightly around the minimum of the potential, we will need to make $ r_{i} $ small. This physically corresponds to having heavy particles in strong harmonic potentials, cf. the definition of the $ r_{i} $. In that case the particles move in such low depths of the potential that they can not perceive the cosine-behaviour of the potential, but simply see a harmonic potential with a slight anharmonicity. This is why the Fock basis works well for the system. More exactly the qubit transition frequencies will be much larger than the coupling strengths and so the number of excitations in the Fock basis will be approximately preserved, again by a rotating wave-type argument. The $ r_{i} $ are exactly the parameters that define the sizes of the $ \phi_{i} $, and so an expansion in $ \phi_{i} $ will in fact be an expansion in $ r_{i} $. The transmon regime is exactly when the $ r_{i} $ are so small that an expansion to fourth order is justified. We will in this paper generally include all orders, but it is helpful to consider the $ r_{i} $ as small to make the physics clearer. A usual value of the $ r_{i} $ is $ <1/6 $, and we will generally work only with even powers of them.

Let us write the Hamiltonian in terms of the creation/annihilation operators
\begin{align}
\begin{split}
H &= -\frac{1}{4r^2}E_{C}\left[(a_{1}^{\dagger} - a_{1})^2 + (a_{2}^{\dagger} - a_{2})^2 + (a_{c}^{\dagger} - a_{c})^2\right] \\
&\hspace{11.5pt} - E\left[2\cos\left(r(a_{1}^{\dagger} + a_{1}) - r(a_{2}^{\dagger} + a_{2})\right)\cos\left(\sqrt{2}r(a_{c}^{\dagger} + a_{c})\right) + \cos\left(\sqrt{2}r(a_{1}^{\dagger} + a_{1}) + \sqrt{2}r(a_{2}^{\dagger} + a_{2})\right)\right]
\end{split}
\end{align}
We will now truncate this to the two lowest levels of each mode and represent the result via Pauli matrices. When we simulate the dynamics of the system we will use the full Hamiltonian above and include more than two levels for non-linear oscillator (normally four), in order to check whether we stay in the qubit space, etc. However, deriving the analytical Hamiltonian for two levels, will make it clearer what we can expect from our system and what we might be able to make it do, even though we will never actually use the two level Hamiltonian for any simulation. Before we reduce to two levels, let us introduce the idea of even and odd interactions. These are interactions that come from an even or odd number of creation or annihilation operators. Hence, XX-couplings are odd while ZZ-couplings are even, and when we include more than two levels there will be interactions that exchange 2, 3 or more excitations. We can see that the $ c $-mode only has even interactions, because its only interaction terms come from a cosine which is an even function. 

The truncation is done with the usual method of calculating the matrix representation of an operator $ A $ by
\begin{align}
M(A) = \mat{\mel{0}{A}{0} & \mel{0}{A}{1} \\ \mel{1}{A}{0} & \mel{1}{A}{1}}
\end{align}
The truncation to the two lowest levels is justified by the anharmonicities (although as mentioned in the main text, these do turn out to be comparable in size to the coupling strengths), which we derive in the next section. The kinetic terms of the Hamiltonian are easiest to calculate and yield
\begin{align}
\begin{split}
(H_{\textup{kin}})_{2} &= -\frac{1}{4r^2}E_{C}\left[(a_{1}^{\dagger} - a_{1})^2 + (a_{2}^{\dagger} - a_{2})^2 + (a_{c}^{\dagger} - a_{c})^2\right] \\
&= -\frac{1}{4r^2}E_{C}\left[(\sigma_{1}^{z} - 2) + (\sigma_{1}^{z} - 2) + (\sigma_{1}^{z} - 2)\right] \\
&= -2Er^2\left[(\sigma_{1}^{z} - 2) + (\sigma_{1}^{z} - 2) + (\sigma_{1}^{z} - 2)\right]
\end{split}
\end{align}
where matrix identities attached to numbers and other matrices are suppressed, i.e. any number added to a matrix is implicitly that number times the identities for each of the two level-systems $ I_{1}\otimes I_{2}\otimes I_{c} $, and for example $ \sigma_{1}^{z} $ is actually $ \sigma_{1}^{z}\otimes I_{2} \otimes I_{c} $. To calculate the contributions from the cosines, we first derive the result quoted in the main text \cref{eq:exptwolevel}. We will write the cosines as complex exponential, and therefore need matrix elements of the form
\begin{align}
\mel{n}{e^{ik(a^{\dagger} + a)}}{m}
\end{align}
for some real constant $ k $, and $ n,m = 0,1 $. The operator is in fact a displacement operator, $ D(\xi) = e^{\xi a^{\dagger} - \xi^{*}a} $, with $ \xi = ik $. We know of displacement operators that
\begin{align}
D(\xi)\ket{0} = e^{-\vert\xi\vert^2/2}\sum_{j = 0}^{\infty}\frac{\xi^{j}}{\sqrt{j!}}\ket{j} \label{eq:dispAct}
\end{align}
and we know its commutation relations, one of which is
\begin{align}
D(\xi)a^{\dagger} = (a^{\dagger} - \xi^{*})D(\xi) \label{eq:dispComm}
\end{align}
With this and the definition $ \ket{n} = \frac{1}{\sqrt{n!}}(a^{\dagger})^n\ket{0} $, we can calculate the needed matrix elements. Incidentally, these identities are sufficient to calculate exactly the matrix elements of any product of sines and cosines of linear combinations of flux coordinates, with any number of energy levels. Let us also note that since the operator $ a^{\dagger} + a $ is symmetric, the operator $ D(ik) = e^{ik(a^{\dagger} + a)} $ is also symmetric, i.e. $ \mel{n}{e^{ik(a^{\dagger} + a)}}{m} = \mel{m}{e^{ik(a^{\dagger} + a)}}{n} $. We find
\begin{align}
\begin{split}
\mel{0}{e^{ik(a^{\dagger} + a)}}{0} &= e^{-k^2/2}\\
\mel{1}{e^{ik(a^{\dagger} + a)}}{0} &= ike^{-k^2/2}\\
\mel{1}{e^{ik(a^{\dagger} + a)}}{1} &= (1 - k^2)e^{-k^2/2}
\end{split}
\end{align}
With these we can write the operator $ e^{ik(a^{\dagger} + a)} $ truncated to two levels in terms of Pauli matrices and the identity
\begin{align}
e^{ik(a^{\dagger} + a)} = \left[\left(1 - \frac{k^2}{2}\right) + \frac{k^2}{2}\sigma^{z} + ik\sigma^{x}\right]e^{-k^2/2} \label{eq:exp2}
\end{align}
Let us now rewrite the potential part of our Hamiltonian in terms of exponentials and use the above identity to truncate it to two levels. First we do the following partial calculations, where we use the fact that the different $ a_{i} $ commute to factorize the exponentials
\begin{align}
\begin{split}
\cos\left(k(a_{1}^{\dagger} + a_{1}) \pm k(a_{2}^{\dagger} + a_{2})\right) &= \frac{1}{2}\left(e^{ik(a_{1}^{\dagger} + a_{1})}e^{\pm ik(a_{2}^{\dagger} + a_{2})} + e^{-ik(a_{1}^{\dagger} + a_{1})}e^{\mp ik(a_{2}^{\dagger} + a_{2})}\right) \\
&= \frac{1}{2}\bigg(\left[\left(1 - \frac{k^2}{2}\right) + \frac{k^2}{2}\sigma_{1}^{z} + ik\sigma_{1}^{x}\right]\left[\left(1 - \frac{k^2}{2}\right) + \frac{k^2}{2}\sigma_{2}^{z} \pm ik\sigma_{2}^{x}\right] \\
&\hspace{25pt} + \left[\left(1 - \frac{k^2}{2}\right) + \frac{k^2}{2}\sigma_{1}^{z} - ik\sigma_{1}^{x}\right]\left[\left(1 - \frac{k^2}{2}\right) + \frac{k^2}{2}\sigma_{2}^{z} \mp ik\sigma_{2}^{x}\right]\bigg)e^{-k^2} \\
&= \left(\left[\left(1 - \frac{k^2}{2}\right) + \frac{k^2}{2}\sigma_{1}^{z}\right]\left[\left(1 - \frac{k^2}{2}\right) + \frac{k^2}{2}\sigma_{2}^{z}\right] \mp k^2\sigma_{1}^{x}\sigma_{2}^{x}\right)e^{-k^2} \\
&= \left(\left(1 - \frac{k^2}{2}\right)^2 + \left(1 - \frac{k^2}{2}\right)\frac{k^2}{2}(\sigma_{1}^{z} + \sigma_{2}^{z}) + \frac{k^4}{4}\sigma_{1}^{z}\sigma_{2}^{z} \mp k^2\sigma_{1}^{x}\sigma_{2}^{x}\right)e^{-k^2} \label{eq:cos}
\end{split}\\
\begin{split}
\cos\left(k(a_{c}^{\dagger} + a_{c})\right) &= \frac{1}{2}\left(e^{ik(a_{c}^{\dagger} + a)} + e^{-ik(a_{c}^{\dagger} + a)}\right) \\
&= \left(\left(1 - \frac{k^2}{2}\right) + \frac{k^2}{2}\sigma^{z}\right)e^{-k^2/2}
\end{split}
\end{align} 
We can then rewrite the potential terms 
\begin{align}
\begin{split}
(H_{\textup{pot}})_{2} &= -E\left[2\cos\left(r(a_{1}^{\dagger} + a_{1}) - r(a_{2}^{\dagger} + a_{2})\right)\cos\left(\sqrt{2}r(a_{c}^{\dagger} + a_{c})\right) + \cos\left(\sqrt{2}r(a_{1}^{\dagger} + a_{1}) + \sqrt{2}r(a_{2}^{\dagger} + a_{2})\right)\right] \\
&= -E\bigg[2\left(\left(1 - \frac{r^2}{2}\right)^2 + \left(1 - \frac{r^2}{2}\right)\frac{r^2}{2}(\sigma_{1}^{z} + \sigma_{2}^{z}) + \frac{r^4}{4}\sigma_{1}^{z}\sigma_{2}^{z} + r^2\sigma_{1}^{x}\sigma_{2}^{x}\right)e^{-r^2}\left(\left(1 - r^2\right) + r^2\sigma_{c}^{z}\right)e^{-r^2} \\
&\hspace{30pt} + \left(\left(1 - r^2\right)^2 + \left(1 - r^2\right)r^2(\sigma_{1}^{z} + \sigma_{2}^{z}) + r^4\sigma_{1}^{z}\sigma_{2}^{z} - 2r^2\sigma_{1}^{x}\sigma_{2}^{x}\right)e^{-2r^2}\bigg] \\
&= -E\bigg[2\left(1 - r^2\right)\left(1 - \frac{r^2}{2}\right)^2 + \left(1 - r^2\right)^2 + \left(1 - r^2\right)\left(2 - \frac{r^2}{2}\right)r^2(\sigma_{1}^{z} + \sigma_{2}^{z}) + 2\left(1 - \frac{r^2}{2}\right)^2r^2\sigma_{c}^{z} \\
&\hspace{25pt} + \left(\frac{3}{2} - \frac{r^2}{2}\right)r^4\sigma_{1}^{z}\sigma_{2}^{z} + \left(1 - \frac{r^2}{2}\right)r^4(\sigma_{1}^{z}\sigma_{c}^{z} + \sigma_{2}^{z}\sigma_{c}^{z}) + \frac{r^6}{2}\sigma_{1}^{z}\sigma_{2}^{z}\sigma_{c}^{z} \\
&\hspace{25pt} + 2r^4\sigma_{1}^{x}\sigma_{2}^{x}(\sigma_{c}^{z} - 1)\bigg]e^{-2r^2}
\end{split}
\end{align}
With these expression we can now put together the two level Hamiltonian. Ignoring constant offsets, which do not change the dynamics anyway, we find
\begin{align}
\begin{split}
H_{2} &= -2Er^2\left[\sigma_{1}^{z} + \sigma_{1}^{z} + \sigma_{1}^{z}\right] \\
&\hspace{11.5pt} - E\bigg[\left(1 - r^2\right)\left(2 - \frac{r^2}{2}\right)r^2(\sigma_{1}^{z} + \sigma_{2}^{z}) + 2\left(1 - \frac{r^2}{2}\right)^2r^2\sigma_{c}^{z} \\
&\hspace{25pt} + \left(\frac{3}{2} - \frac{r^2}{2}\right)r^4\sigma_{1}^{z}\sigma_{2}^{z} + \left(1 - \frac{r^2}{2}\right)r^4(\sigma_{1}^{z}\sigma_{c}^{z} + \sigma_{2}^{z}\sigma_{c}^{z}) + \frac{r^6}{2}\sigma_{1}^{z}\sigma_{2}^{z}\sigma_{c}^{z} \\
&\hspace{25pt} + 2r^4\sigma_{1}^{x}\sigma_{2}^{x}(\sigma_{c}^{z} - 1)\bigg]e^{-2r^2} \\
&= -E\left(2r^2 + \frac{1}{2}\left(1 - r^2\right)\left(4 - r^2\right)r^2e^{-2r^2}\right)(\sigma_{1}^{z} + \sigma_{2}^{z}) - E\left(2r^2 + \frac{1}{2}\left(2 - r^2\right)^2r^2e^{-2r^2}\right)\sigma_{c}^{z} \\
&\hspace{11.5pt} - \frac{3}{2}Er^4e^{-2r^2}\sigma_{1}^{z}\sigma_{2}^{z} - E\left(1 - \frac{r^2}{2}\right)r^4e^{-2r^2}(\sigma_{1}^{z}\sigma_{c}^{z} + \sigma_{2}^{z}\sigma_{c}^{z}) \\
&\hspace{11.5pt} + Er^6e^{-2r^2}\sigma_{1}^{z}\sigma_{2}^{z}\frac{1 - \sigma_{c}^{z}}{2} \\
&\hspace{11.5pt} + 4Er^4e^{-2r^2}\sigma_{1}^{x}\sigma_{2}^{x}\frac{1 - \sigma_{c}^{z}}{2}
\end{split}
\end{align}
Defining the following constants
\begin{align}
\begin{split}
\Omega_{q} &= E\left(4r^2 + \left(1 - r^2\right)\left(4 - r^2\right)r^2e^{-2r^2}\right)\\
\Omega_{c} &= E\left(4r^2 + \left(2 - r^2\right)^2r^2e^{-2r^2}\right)\\
J_{qq}^{z} &= \frac{3}{2}Er^4e^{-2r^2}\\
J_{qc}^{z} &= E\left(1 - \frac{r^2}{2}\right)r^4e^{-2r^2}\\
J_{qqc}^{z} &= Er^6e^{-2r^2}\\
J_{qqc}^{x} &= 4Er^4e^{-2r^2}
\end{split}
\end{align}
we get the Hamiltonian
\begin{align}
\begin{split}
H_{2} &= -\frac{1}{2}\Omega_{q}(\sigma_{1}^{z} + \sigma_{2}^{z}) - \frac{1}{2}\Omega_{c}\sigma_{c}^{z}\\
&\hspace{11.5pt} - J_{qq}^{z}\sigma_{1}^{z}\sigma_{2}^{z} - J_{qc}^{z}(\sigma_{1}^{z}\sigma_{c}^{z} + \sigma_{2}^{z}\sigma_{c}^{z}) + J_{qqc}^{z}\sigma_{1}^{z}\sigma_{2}^{z}\frac{1 - \sigma_{c}^{z}}{2}\\
&\hspace{11.5pt} + J_{qqc}^{x}\sigma_{1}^{x}\sigma_{2}^{x}\frac{1 - \sigma_{c}^{z}}{2}
\end{split}
\end{align}
as quoted in the main text. We see that there are no XX-type couplings other than the desired XXC-coupling. The XXC- and ZZC-couplings could of course both be written as two independent couplings, a two-particle and a three-particle interaction. Writing the XX-type couplings as a single XXC-coupling, however, makes it clear that they disappear completely when the $ c $-qubit is in the excited state. The XX-couplings are controlled by the state of the $ c $-qubit. For the ZZC-coupling it is a bit more arbitrary, as there is an additional ZZ-coupling between qubits $ 1 $ and $ 2 $, i.e. even if we have the $ c $-qubit in its ground state there is still a residual ZZ-coupling. 

The gate operation time, $ T_{g} $, defined as the time it takes for $ \ket{0_{1}1_{2}1_{c}} $ to evolve to $ \ket{1_{1}0_{2}1_{c}} $ and vice versa, can be easily calculated from considering a two level system. If we have two energy levels detuned by $ \Delta $ interacting with a coupling strength $ J $, described by the Hamiltonian
\begin{align}
H_{\textup{two level}} = \mat{\frac{1}{2}\Delta & J \\ J & -\frac{1}{2}\Delta}
\end{align}
and initialise in one of the states, then it will evolve into the other in a time
\begin{align}
T_{g} = \frac{\pi}{2\sqrt{\Delta^2/4 + J^2}}
\end{align} 
with a maximal amplitude
\begin{align}
A = \frac{J^2}{J^2 + \Delta^2/4}e^{i\frac{\pi}{2}\sqrt{\frac{\Delta^2}{4J^2} + 1}}
\end{align}
Hence, as $ \ket{0_{1}1_{2}1_{c}} $ and $ \ket{1_{1}0_{2}1_{c}} $ can be considered two resonant levels ($ \Delta = 0 $) that interact via $ J = J_{12c}^{x} $, we would expect the gate times to be given by the above formula, and the transfer to happen with an amplitude of $ e^{i\frac{\pi}{2}} $, i.e. complete transfer of population with a phase of $ \frac{\pi}{2} $. Of course for a realistic system noise will decrease the amplitude and give it a different phase. Furthermore, the non-zero XX-coupling we find in the more realistic version of our circuit and higher order interactions induced by the strong couplings to higher levels will change the gate time and transfer amplitude. 

\subsection{Anharmonicities}\label{app:idealanharm}
We can calculate the anharmonicities exactly using the same method as above. The anharmonicity is defined as
\begin{align}
\alpha = (\mel{2}{H}{2} - \mel{1}{H}{1}) - (\mel{1}{H}{1} - \mel{0}{H}{0}) = \mel{2}{H}{2} + \mel{0}{H}{0} - 2\mel{1}{H}{1}
\end{align}
As there are multiple qubits in play here, there will be some contributions to the anharmonicities that come from interactions and are as such dependent on the state of the qubits. This is similar to how ZZ-couplings shift the energy levels depending on the states of the qubits. These contributions we ignore in the above definition, i.e. for this calculation we will only consider terms in the Hamiltonian that are not interactions. The contributions from interactions will always be smaller, as they are at least a factor $ r^2 $ smaller, and the non-interaction contribution are themselves at least of order $ r^4 $. Remember that $ r $ is the small parameter and usually one expands to only fourth order, thus interaction contributions to the anharmonicities are usually ignored as well. In addition to the matrix elements found before, we will need
\begin{align}
\mel{2}{e^{ik(a^{\dagger} + a)}}{2} = \left(1 - 2k^2 + \frac{k^4}{2}\right)e^{-k^2/2}
\end{align}
To make sure we do not include contributions that come from ZZ-like couplings, we shall write the diagonal part of $ e^{ik(a^{\dagger} + a)} $ as a 3x3 matrix in terms of the identity and the following two matrices
\begin{align}
Z &= \mat{1 & 0 & 0 \\ 0 & -1 & 0 \\ 0 & 0 & -3}\\
A &= \mat{0 & 0 & 0 \\ 0 & 0 & 0 \\ 0 & 0 & 1}
\end{align}
The first is continuation of $ \sigma^{z} $ and the second is the matrix that will carry the contribution to the anharmonicity (both the identity and $ Z $ do not contribute to the anharmonicity). With $ \mel{2}{e^{ik(a^{\dagger} + a)}}{2} $ calculated we can see that the diagonal part of $ e^{ik(a^{\dagger} + a)} $ can be written as $ \left(\left(1 - \frac{k^2}{2}\right) + \frac{k^2}{2}Z + \frac{k^4}{2}A\right)e^{-k^2/2} $, which is consistent with the two level form of $ e^{ik(a^{\dagger} + a)} $ we found previously. The idea is now that when calculating for example $ a_{1} $, and faced with product of exponentials, like $ e^{ik(a_{1}^{\dagger} + a_{1})}e^{ik(a_{2}^{\dagger} + a_{2})} $, we shall take the second exponential to be $ \left(1 - \frac{k^2}{2}\right) $ times the identity, rather than for example just the identity alone. This factor is a result of the basis of operators we use (particularly the use of $ \sigma^{z} $). For a different basis the anharmonicity would be differently distributed between interaction and non-interaction contributions. Let us finally note that we must necessarily have $ \alpha_{1} = \alpha_{2} $ because the Hamiltonian is symmetric under the exchange of the indices $ 1 $ and $ 2 $. Replacing exponentials with $ \left(\left(1 - \frac{k^2}{2}\right) + \frac{k^2}{2}Z + \frac{k^4}{2}A\right)e^{-k^2/2} $ and expanding all products, we can simply sum the coefficients of the stand-alone $ A $'s for each mode. Doing this we get
\begin{align}
\begin{split}
\alpha_{1} = \alpha_{2} &= -E\left(3 - \frac{r^2}{2}\right)\left(1-r^2\right)r^4e^{-2r^2}\\
\alpha_{c} &= -E\left(2 - r^2\right)^2r^4e^{-2r^2}
\end{split}
\end{align}

We can note here that $ Er^4 = \frac{E_{C}}{8} $, revealing that the anharmonicities go as the charging energies, as usual with transmon qubits. Furthermore, we can see that the anharmonicities are very much comparable to the ZZ- and XXC-coupling strengths found above. Indeed, to fourth order in $ r $, we have $ \alpha_{1} = \alpha_{2} = \frac{3}{4}\alpha_{c} $ and $ \alpha_{c} = J_{qqc}^{x} = \frac{8}{3}J_{qq}^{z} = 4J_{qc}^{z} $. As mentioned in the main text, this means we cannot make use of all the available states in the qubit subspace, i.e. the space of just $ 0 $ or $ 1 $ excitation in each mode, as some of them couple strongly outside of the qubit subspace.

\subsection{Gate}\label{app:idealgate}
As the anharmonicities are not large enough to suppress interactions with higher levels we must restrict the available states further than just the qubit subspace. In particular the two states $ \ket{1_{1}1_{2}0_{c}} $ and $ \ket{1_{1}1_{2}1_{c}} $ with an excitation in both of qubits $ 1 $ and $ 2 $ couples non-negligibly to states out of the qubit subspace. In these states the excitations in qubits $ 1 $ and $ 2 $ can come together in just one of the modes, exciting it out of the qubits subspace. Similarly these two excitations can even go into the $ c $-mode, via an XXZ-like interaction, because the $ c $-mode is close to being resonant with the other two qubits. All other states, however, do not evolve out of the qubit subspace. In these states there are at most two excitations available and in that case one of them will be in the $ c $-mode. As it is only coupled via even interactions this one excitation cannot leave the $ c $-mode and so no mode can acquire both excitations.

This means that as long as we do not initialize the system with more than one excitation in total in qubits $ 1 $ and $ 2 $, we expect the state of the $ c $-mode to be stationary. Hence, whether the XXC-coupling is turned on or off is completely determined by initialization and does not change during dynamics, barring noise effects. So working in the basis of $ \ket{0_{1}0_{2}\phi_{c}} $, $ \ket{1_{1}0_{2}\phi_{c}} $ and $ \ket{0_{1}1_{2}\phi_{c}} $ with $ \phi_{c} = 0,1 $, we may say that the system implements a gate represented by a diagonal matrix of phases when $ \phi_{c} = 0 $ and by
\begin{align}
U = \mat{e^{i\gamma} & 0 & 0 \\ 0 & 0 & e^{i\delta} \\ 0 & e^{i\delta} & 0}
\end{align}
when $ \phi_{c} = 1 $. The phases are a simple consequence of the fact that the different states have different energies, and so acquire different trivial dynamical phases. As mentioned in the main text, this gate could be used as a CNOT with $ \ket{1_{1}0_{2}} $ and $ \ket{0_{1}1_{2}} $ as the logical $ \ket{0} $ and $ \ket{1} $ states and the $ c $-qubit as control. Naming one of the qubits $ 1 $ and $ 2 $ as input and the other as output, and always initializing with the output in the ground state, the gate can also implement a spin transistor that allows the transport of some superposition state from input to output conditioned on the state of the $ c $-qubit.

\subsection{External coupling}\label{app:readout}
Although we are performing a change of coordinates we can still couple each of the new coordinates separately for external control. If we couple capacitively to each of the original node fluxes, the correspond terms in the Hamiltonian take the form
\begin{align}
\frac{K_{j}}{2}(\dot{\psi}_{j} - D_{j})^2
\end{align}
for $ j = 1,2,3,4 $, where $ K_{j} $ are the capacitances of the couplings and $ D_{j} $ are drives, i.e. essentially fluxes that we can control externally. Writing these capacitances in a diagonal matrix $ \mathcal{K}_{D} $ and the drives in a vector $ \textbf{D}_{\psi} $, we can write all the new terms as
\begin{align}
H_{D} = \frac{1}{2}\dot{\bm{\psi}}^{T}\mathcal{K}_{D}\dot{\bm{\psi}} - \textbf{D}_{\psi}^{T}\mathcal{K}_{D}\dot{\bm{\psi}}
\end{align}
where we have ignored the terms that do not contain any node fluxes. After the coordinate transformation, $ \bm{\psi} = V\bm{\phi} $, we have
\begin{align}
H_{D} = \frac{1}{2}\dot{\bm{\phi}}^{T}V^{T}\mathcal{K}_{D}V\dot{\bm{\phi}} - \textbf{D}_{\psi}^{T}\mathcal{K}_{D}V\dot{\bm{\phi}}
\end{align}
The first term will induce undesired couplings between the modes if $ V^{T}\mathcal{K}_{D}V $ is not diagonal. Calculations show that this happens only when $ \mathcal{K}_{D} $ has identical entries on its diagonal, i.e. all couplings to external control must be identical. 

As for the second term, we would want it to be on the same form as before the coordinate transformation, such that we have a drive for each mode coupling to only that mode. If we introduce new drives $ \textbf{D}_{\phi} $, and write $ \textbf{D}_{\psi} = W\textbf{D}_{\phi} $ for some orthogonal matrix W, the second term in $ H_{D} $ becomes 
\begin{align}
-\textbf{D}_{\phi}^{T}W^{T}\mathcal{K}_{D}V\dot{\bm{\phi}}
\end{align}
We thus want $ W^{T}\mathcal{K}_{D}V $ to be equal to some diagonal matrix $ \mathcal{C}_{D} $ with unit of capacitance, which it will be if we choose
\begin{align}
W = \left(\mathcal{C}_{D}(\mathcal{K}_{D}V)^{-1}\right)^{T} = \mathcal{K}_{D}^{-1}V\mathcal{C}_{D}
\end{align}
We have here used that $ V $ is orthogonal and both $ \mathcal{K}_{D} $ and $ \mathcal{C}_{D} $ are diagonal, which also means that $ \mathcal{K}_{D} $ is trivially invertible. Since the entries of $ \mathcal{K}_{D} $ ideally are identical, we can simply choose $ W = V $ (i.e. $ \mathcal{C}_{D} = \mathcal{K}_{D} $). Hence, identical couplings to the original coordinates with drives $ \textbf{D}_{\psi} = V\textbf{D}_{\phi} $, will result in identical couplings between the new coordinates $ \bm{\phi} $ and $ \textbf{D}_{\phi} $. If the couplings are not identical we can still talk to each mode directly, but there will be induced XX-couplings between them. These will be small though, for small discrepancies.

\subsection{Realistic circuit}\label{app:realcirc}
Let us now go through all the same quantities but for the more realistic circuit, where we have included capacitive couplings to external control and readout and to ground. This circuit can be seen in \cref{fig:circuit} of the main text, where the loose ends pointing up collectively represent the external couplings to control, readout and ground. We assume that these are identical for each node, and contribute a capacitance $ K $ to each entry in the diagonal of the capacitance matrix. As mentioned this introduces an asymmetry between the modes, as we will go through in a moment, which we counter by adding a new branch to the circuit, connecting its two ends, consisting of an inductor with inductance $ L_{0} $ and a parasitic capacitance $ C_{0} $. Furthermore to cancel interaction terms from this inductor we add an identical one to the middle block of the original circuit. To retain the same modes of the capacitance network, we must also add $ C_{0} $ to the capacitance of the middle block. The capacitance matrix becomes
\begin{align}
\mathcal{K} = \mat{C+C_{0}+K & -C & 0 & -C_{0} \\ -C & 2C+C_{0}+K & -C-C_{0} & 0 \\ 0 & -C-C_{0} & 2C+C_{0}+K & -C \\ -C_{0} & 0 & -C & C+C_{0}+K}
\end{align}
This matrix has the same eigenvectors as before, i.e. the capacitance network has the same modes. If we change basis to $ \bm{\phi} $, we get
\begin{align}
\mathcal{C} = \mat{4K & 0 & 0 & 0 \\ 0 & 8C + 4(2 + \sqrt{2})(2C_{0} + K) & 0 & 0 \\ 0 & 0 & 8C + 4(2 - \sqrt{2})(2C_{0} + K) & 0 \\ 0 & 0 & 0 & 8C + 4K}
\end{align}
The center of mass-mode has now acquired a non-zero entry, but it still does not interact with any other mode, so we will still be ignoring it. More importantly the entries pertaining to $ \phi_{1} $ and $ \phi_{2} $ are no longer identical, even for $ C_{0} = 0 $. This is the asymmetry that first and foremost creates a detuning between the modes and also makes the uncontrolled XX-coupling not cancel. Calculations show that even for $ C_{0} = 0 $ and $ K = \SI{1}{\femto\farad} $ qubits $ 1 $ and $ 2 $ are significantly detuned. This asymmetry, particularly the detuning, is countered by tuning $ L_{0} $.

The kinetic part of the Hamiltonian can again be written in terms of the conjugate momenta
\begin{align}
H_{\textup{kin}} = \frac{1}{2}\textbf{p}^{T}\mathcal{C}^{-1}\textbf{p} = \frac{\mathcal{C}_{1}^{-1}}{2}p_{1}^2 + \frac{\mathcal{C}_{2}^{-1}}{2}p_{2}^2 + \frac{\mathcal{C}_{c}^{-1}}{2}p_{c}^2
\end{align}
where we have again removed $ \phi_{CM} $-entries from the vectors and matrices. The potential part is 
\begin{align}
\begin{split}
H_{\textup{pot}} &= -E\cos(\sqrt{2}\phi_{1} - \sqrt{2}\phi_{2} + 2\phi_{c}) - E\cos(2\phi_{1} + 2\phi_{2}) + E_{L_{0}}(2\phi_{1} + 2\phi_{2})^2\\
&\hspace{11.5pt} - E\cos(\sqrt{2}\phi_{1} - \sqrt{2}\phi_{2} - 2\phi_{c}) + E_{L_{0}}(2(\sqrt{2} + 1)\phi_{1} - 2(\sqrt{2} - 1)\phi_{2})^2\\
&= -E\left[2\cos\left(\sqrt{2}\phi_{1} - \sqrt{2}\phi_{2}\right)\cos(2\phi_{c}) + \cos(2\phi_{1} + 2\phi_{2})\right]\\
&\hspace{11.5pt} + 8E_{L_{0}}\left[(2+\sqrt{2})\phi_{1}^2 + (2-\sqrt{2})\phi_{2}^2\right]
\end{split}
\end{align}
where we have introduced the inductive energy $ E_{L_{0}} = \frac{1}{2L_{0}} $. Notice how the inductor terms are not symmetric under the exchange of the indices $ 1 $ and $ 2 $. This is the asymmetry we will use to counter the asymmetry of the capacitance matrix. As before we can now extract the $ \phi_{i}^2 $-terms and define the $ r_{i} $. Doing so we find
\begin{align}
\begin{split}
r_{1} &= \left(\frac{E_{C}E_{C_{0}}E_{K}}{4(2E_{C_{0}}E_{K} + (2+\sqrt{2})(2E_{C}E_{K} + E_{C}E_{C_{0}}))(E + 2(2+\sqrt{2})E_{L_{0}})}\right)^{1/4}\\
r_{2} &= \left(\frac{E_{C}E_{C_{0}}E_{K}}{4(2E_{C_{0}}E_{K} + (2-\sqrt{2})(2E_{C}E_{K} + E_{C}E_{C_{0}}))(E + 2(2-\sqrt{2})E_{L_{0}})}\right)^{1/4}\\
r_{c} &= \left(\frac{E_{C}E_{K}}{4(2E_{K} + E_{C})E}\right)^{1/4}
\end{split}
\end{align}
These expressions indicate that the asymmetry between $ \phi_{1} $ and $ \phi_{2} $ may be contained in the sign of the factor $ 2 \pm \sqrt{2} $. Furthermore it seems justified that the asymmetry can be quantified by the ratios
\begin{align}
\frac{2E_{C}E_{K} + E_{C}E_{C_{0}}}{2E_{C_{0}}E_{K}} = \frac{2C_{0} + K}{2C}, \quad \text{ and } \quad \frac{2E_{L_{0}}}{E}
\end{align}
for capacitive and inductive contributions respectively. As we would like the asymmetries to cancel each other, we expect that these ratios have to be comparable in size.

We promote $ \phi_{i} $ and $ p_{i} $ to operators with commutation relation $ [\phi_{i},p_{i}] = i $, and we introduce creation/annihilation operators $ a_{i}^{\dagger} $/$ a_{i} $ for each mode, so that we may write
\begin{align}
\phi_{i}  = \frac{r_{i}}{\sqrt{2}}(a_{i}^{\dagger} + a_{i}), \qquad p_{i} = i\frac{1}{\sqrt{2}r_{i}}(a_{i}^{\dagger} - a_{i})
\end{align}
We can then write the Hamiltonian in terms of creation/annihilation operators
\begin{align}
\begin{split}
H &= -2(E + 2(2+\sqrt{2})E_{L_{0}})r_{1}^2(a_{1}^{\dagger} - a_{1})^2 - 2(E + 2(2-\sqrt{2})E_{L_{0}})r_{2}^2(a_{2}^{\dagger} - a_{2})^2 - 2Er_{c}^2(a_{c}^{\dagger} - a_{c})^2 \\
&\hspace{11.5pt} - E\left[2\cos\left(r_{1}(a_{1}^{\dagger} + a_{1}) - r_{2}(a_{2}^{\dagger} + a_{2})\right)\cos(\sqrt{2}r_{c}(a_{c}^{\dagger} + a_{c})) + \cos(\sqrt{2}r_{1}(a_{1}^{\dagger} + a_{1}) + \sqrt{2}r_{2}(a_{2}^{\dagger} + a_{2}))\right] \\
&\hspace{11.5pt}  + 4E_{L_{0}}\left[(2+\sqrt{2})r_{1}^2(a_{1}^{\dagger} + a_{1})^2 + (2-\sqrt{2})r_{2}^2(a_{2}^{\dagger} + a_{2})^2\right]
\end{split}
\end{align}
Again, the asymmetry between $ \phi_{1} $ and $ \phi_{2} $ appears to be contained in the sign $ 2 \pm \sqrt{2} $. As before we can use \cref{eq:exp2} to truncate the cosines to two levels. Let us first note a generalization of \cref{eq:cos}
\begin{align}
\begin{split}
\cos&\left(k_{1}(a_{1}^{\dagger} + a_{1}) + k_{2}(a_{2}^{\dagger} + a_{2})\right)\\ 
&= \left(\left(1 - \frac{k_{1}^2}{2}\right)\left(1 - \frac{k_{2}^2}{2}\right) + \left(1 - \frac{k_{2}^2}{2}\right)\frac{k_{1}^2}{2}\sigma_{1}^{z} + \left(1 - \frac{k_{1}^2}{2}\right)\frac{k_{2}^2}{2}\sigma_{2}^{z} + \frac{k_{1}^2k_{2}^2}{4}\sigma_{1}^{z}\sigma_{2}^{z} - k_{1}k_{2}\sigma_{1}^{x}\sigma_{2}^{x}\right)e^{-(k_{1}^2+k_{2}^2)/2}
\end{split}
\end{align}
We find a Hamiltonian of the form
\begin{align}
\begin{split}
H_{2} &= -\frac{1}{2}\Omega_{1}\sigma_{1}^{z} - \frac{1}{2}\Omega_{2}\sigma_{2}^{z} - \frac{1}{2}\Omega_{c}\sigma_{c}^{z} \\
&\hspace{11.5pt} - J_{12}^{z}\sigma_{1}^{z}\sigma_{2}^{z} - J_{1c}^{z}\sigma_{1}^{z}\sigma_{c}^{z} - J_{2c}^{z}\sigma_{2}^{z}\sigma_{c}^{z} + J_{12c}^{z}\sigma_{1}^{z}\sigma_{2}^{z}\frac{1 - \sigma_{c}^{z}}{2} \\
&\hspace{11.5pt} + J_{12}^{x}\sigma_{1}^{x}\sigma_{2}^{x} + J_{12c}^{x}\sigma_{1}^{x}\sigma_{2}^{x}\frac{1 - \sigma_{c}^{z}}{2}
\end{split}
\end{align}
with
\begin{align}
\begin{split}
\Omega_{1} &= 2E\left[2r_{1}^2 + r_{1}^2\left(1 - \frac{r_{2}^2}{2}\right)\left(1 - r_{c}^2\right)e^{-(r_{1}^2+r_{2}^2 + 2r_{c}^2)/2} + r_{1}^2\left(1 - r_{2}^2\right)e^{-(r_{1}^2+r_{2}^2)}\right]\\
&\hspace{11.5pt} + 16(2+\sqrt{2})E_{L_{0}}r_{1}^2\\
\Omega_{2} &= 2E\left[2r_{2}^2 + \left(1 - \frac{r_{1}^2}{2}\right)r_{2}^2\left(1 - r_{c}^2\right)e^{-(r_{1}^2+r_{2}^2 + 2r_{c}^2)/2} + \left(1 - r_{1}^2\right)r_{2}^2e^{-(r_{1}^2+r_{2}^2)}\right]\\
&\hspace{11.5pt} + 16(2-\sqrt{2})E_{L_{0}}r_{2}^2\\
\Omega_{c} &= 4E\left[r_{c}^2 + \left(1 - \frac{r_{1}^2}{2}\right)\left(1 - \frac{r_{2}^2}{2}\right)r_{c}^2e^{-(r_{1}^2+r_{2}^2 + 2r_{c}^2)/2}\right]\\
J_{12}^{z} &= E\left[\frac{1}{2}e^{-(r_{1}^2+r_{2}^2 + 2r_{c}^2)/2} + e^{-(r_{1}^2+r_{2}^2)}\right]r_{1}^2r_{2}^2\\
J_{1c}^{z} &= Er_{1}^2\left(1 - \frac{r_{2}^2}{2}\right)r_{c}^2e^{-(r_{1}^2+r_{2}^2 + 2r_{c}^2)/2}\\
J_{2c}^{z} &= E\left(1 - \frac{r_{1}^2}{2}\right)r_{2}^2r_{c}^2e^{-(r_{1}^2+r_{2}^2 + 2r_{c}^2)/2}\\
J_{12}^{x} &= 2E\left[1 - e^{-(2r_{c}^2 - r_{1}^2 - r_{2}^2)/2}\right]e^{-(r_{1}^2+r_{2}^2)}r_{1}r_{2}\\
J_{12c}^{z} &= Er_{1}^2r_{2}^2r_{c}^2e^{-(r_{1}^2+r_{2}^2 + 2r_{c}^2)/2}\\
J_{12c}^{x} &= 4Er_{1}r_{2}r_{c}^2e^{-(r_{1}^2+r_{2}^2 + 2r_{c}^2)/2}
\end{split}
\end{align}
Hence, we have three qubits, with generally different transition frequencies, interacting via ZZ-, XX-, ZZC- and XXC-couplings. The Hamiltonian has the same form as in the idealized circuit except for the XX-coupling and the fact that qubits $ 1 $ and $ 2 $ are not necessarily resonant. Likewise we can calculate the anharmonicities and find
\begin{align}
\begin{split}
\alpha_{1} &= -E\left[\left(1 - \frac{r_{2}^2}{2}\right)\left(1 - r_{c}^2\right)e^{-(r_{1}^2 + r_{2}^2 + 2r_{c}^2)/2} + 2\left(1 - r_{2}^2\right)e^{-(r_{1}^2 + r_{2}^2)}\right]r_{1}^4\\
\alpha_{2} &= -E\left[\left(1 - \frac{r_{1}^2}{2}\right)\left(1 - r_{c}^2\right)e^{-(r_{1}^2 + r_{2}^2 + 2r_{c}^2)/2} + 2\left(1 - r_{1}^2\right)e^{-(r_{1}^2 + r_{2}^2)}\right]r_{2}^4\\
\alpha_{c} &= -E\left(2 - r_{1}^2\right)\left(2 - r_{2}^2\right)r_{c}^4e^{-(r_{1}^2 + r_{2}^2 + 2r_{c}^2)/2}
\end{split}
\end{align}


\subsection{Four level exact exponential operator}\label{app:fourlevel}
We use the method of \cref{eq:dispAct,eq:dispComm} to calculate a four level version of \cref{eq:exp2}. First we need more matrix elements. Supplementing our previous list we find
\begin{align}
\begin{split}
\mel{0}{e^{ik(a^{\dagger} + a)}}{0} &= e^{-k^2/2}\\
\mel{1}{e^{ik(a^{\dagger} + a)}}{0} &= ike^{-k^2/2}\\
\mel{2}{e^{ik(a^{\dagger} + a)}}{0} &= -\frac{k^2}{\sqrt{2}}e^{-k^2/2}\\
\mel{3}{e^{ik(a^{\dagger} + a)}}{0} &= -\frac{ik^3}{\sqrt{6}}e^{-k^2/2}\\
\mel{1}{e^{ik(a^{\dagger} + a)}}{1} &= (1 - k^2)e^{-k^2/2}\\
\mel{2}{e^{ik(a^{\dagger} + a)}}{1} &= \sqrt{2}ik\left(1 - \frac{k^2}{2}\right)e^{-k^2/2}\\
\mel{3}{e^{ik(a^{\dagger} + a)}}{1} &= -\sqrt{6}\frac{k^2}{2}\left(1 - \frac{k^2}{3}\right)e^{-k^2/2}\\
\mel{2}{e^{ik(a^{\dagger} + a)}}{2} &= \left(1 - 2k^2 + \frac{k^4}{2}\right)e^{-k^2/2}\\
\mel{3}{e^{ik(a^{\dagger} + a)}}{2} &= \sqrt{3}ik\left(1 - k^2 + \frac{k^4}{6}\right)e^{-ik^2/2}\\
\mel{3}{e^{ik(a^{\dagger} + a)}}{3} &= \left(1 - 3k^2 + \frac{3k^4}{2} - \frac{k^6}{6}\right)e^{-ik^2/2}
\end{split}
\end{align}
We introduce the matrices
\begin{align}
\begin{split}
Z &= \mat{1 & 0 & 0 & 0 \\ 0 & -1 & 0 & 0 \\ 0 & 0 & -3 & 0 \\ 0 & 0 & 0 & -5}\\
A &= \mat{0 & 0 & 0 & 0 \\ 0 & 0 & 0 & 0 \\ 0 & 0 & 1 & 0 \\ 0 & 0 & 0 & 3}\\
B &= \mat{0 & 0 & 0 & 0 \\ 0 & 0 & 0 & 0 \\ 0 & 0 & 0 & 0 \\ 0 & 0 & 0 & 1}
\end{split}
\end{align}
and finally $ X_{ij} $ for $ i,j = 0,1,2,3 $ and $ i<j $, whose $ a,b $'th entries are
\begin{align}
(X_{ij})_{ab} &= \begin{cases}
0, & \textup{ for } ab \neq ij,ji\\
1, & \textup{ for } ab = ij,ji
\end{cases}
\end{align}
Together with the identity $ Z $, $ A $ and $ B $ describe the contributions to the energy levels. In particular $ Z $ is similar to $ \sigma^{z} $, while $ A $ describes the anharmonicity and $ B $ describes a similar anharmonic energy shift beyond the regular anharmonicity only relevant for the third excited state and higher. In terms of the usual bosonic number operator $ N = a^{\dagger}a = \diag{0,1,2,3} $, we may say that $ Z $ corresponds to $ N $, $ A $ corresponds to $ N^2 $ and $ B $ to $ N^3 $. Alternatively, we may say that $ A $ corresponds to $ a^{\dagger}a^{\dagger}aa $ and $ B $ to $ a^{\dagger}a^{\dagger}a^{\dagger}aaa $, which shows how $ A $ does not affect levels below the second excited one, while $ B $ does not matter for levels below the third excited. The $ X_{ij} $ describe rotation or flipping between the $ i $'th and $ j $'th energy level, just like $ \sigma^{x} $ describes rotation or flipping between the ground and excited state in the two-level case. In terms of these matrices, we can write
\begin{align}
\begin{split}
\left(e^{ik(a^{\dagger} + a)}\right)_{4} &= \bigg[\left(1 - \frac{k^2}{2}\right) + \frac{k^2}{2}Z + \frac{k^4}{2}A - \frac{k^6}{6}B \\
&\hspace{11.5pt} + ikX_{01} - \frac{k^2}{\sqrt{2}}X_{02} - \frac{ik^3}{\sqrt{6}}X_{03} + \sqrt{2}ik\left(1 - \frac{k^2}{2}\right)X_{12} \\
&\hspace{11.5pt} - \sqrt{6}\frac{k^2}{2}\left(1 - \frac{k^2}{3}\right)X_{13} + \sqrt{3}ik\left(1 - k^2 + \frac{k^4}{6}\right)X_{23}\bigg]e^{-k^2/2}
\end{split}
\end{align}
To reduce this to the two-level version one must merely exchange $ Z $ and $ X_{01} $ with $ \sigma^{z} $ and $ \sigma^{x} $ respectively, and remove all other matrices except the identity.

\end{document}